\documentclass[a4paper]{aa}
\usepackage{txfonts,graphicx,natbib,aalongtable}
\bibpunct{(}{)}{;}{a}{}{,}
\newcommand{\cz}{\ensuremath{C_Z}}
\newcommand{\pv}{\ensuremath{P_V}}
\newcommand{\nv}{\ensuremath{N_V}}
\newcommand{\pvr}{\ensuremath{P_V^\mathrm{(rect)}}}
\newcommand{\nvr}{\ensuremath{N_V^\mathrm{(rect)}}}
\newcommand{\bz}{\ensuremath{\langle B_z\rangle}}
\newcommand{\sbz}{\ensuremath{\sigma_{\langle B_z\rangle}}}
\newcommand{\nz}{\ensuremath{\langle N_z\rangle}}
\newcommand{\snz}{\ensuremath{\sigma_{\langle N_z\rangle}}}
\newcommand{\fo}{\ensuremath{f^\parallel}}
\newcommand{\fe}{\ensuremath{f^\perp}}
\newcommand{\dhh}{\ensuremath{\delta h}}
\newcommand{\alphao}{\ensuremath{\alpha_0}}
\newcommand{\dalphao}{\ensuremath{\Delta A}}
\newcommand{\kms}{km\,s\ensuremath{^{-1}}}
\begin{document}
\title{Magnetic field measurements and their uncertainties:\\ the FORS1 legacy}
       \author{
        S.~Bagnulo      \inst{1}
       \and
        J.D.~Landstreet  \inst{1,2}
       \and
        L.~Fossati      \inst{3}
       \and
        O.~Kochukhov    \inst{4}
        }
\institute{
           Armagh Observatory,
           College Hill,
           Armagh BT61 9DG,
           Northern Ireland, U.K.
           \email{sba@arm.ac.uk}
           \and
           Physics \& Astronomy Department,
           The University of Western Ontario,
           London, Ontario, Canada N6A 3K7.\\
           \email{jlandstr@uwo.ca}
           \and
           Department of Physics and Astronomy,
           Open University, Walton Hall,
           Milton Keynes MK7 6AA, UK.\\
           \email{l.fossati@open.ac.uk}
           \and
           Department of Physics and Astronomy,
           Uppsala University,
           751 20 Uppsala, Sweden.
           \email{Oleg.Kochukhov@fysast.uu.se}
}

\authorrunning{S.\ Bagnulo et al.}
\titlerunning{Magnetic field measurements and their uncertainties: the FORS1 legacy}

\date{Received: 2011-09-15 / Accepted: 2011-11-24}
\abstract
% Context
{
During the last decade, the FORS1 instrument of the ESO Very Large
Telescope has been extensively used to study stellar magnetism. A
number of interesting discoveries of magnetic fields in several
classes of stars have been announced, many of which obtained at a $\sim
3\,\sigma$ level; some of the discoveries are confirmed by
measurements obtained with other instruments, some are not.
}
%Aims
{
We investigate the reasons for the discrepancies between the results
obtained with FORS1 and those obtained with other instruments.
}
%Method
{ 
Using the ESO FORS pipeline, we have developed a semi-automatic
procedure for magnetic field determination. We have applied this
procedure to the full content of circular spectropolarimetric
measurements of the FORS1 archive (except for most of the observations
obtained in multi-object spectropolarimetric mode). We have devised and
applied a number of consistency checks to our field determinations, and we
have compared our results to those previously published in the literature.
}
%Results
{
We find that for high signal-to-noise ratio measurements, photon noise
does not account for the full error bars. We discuss how field
measurements depend on the specific algorithm adopted for data
reduction, and we show that very small instrument flexures, negligible
in most of the instrument applications, may be responsible for some
spurious field detections in the null profiles. Finally, we find
that we are unable to reproduce some results previously published
in the literature. Consequently, we do not confirm some important
discoveries of magnetic fields obtained with FORS1 and reported
in previous publications.
}
%Conclusions
{
Our revised field measurements show that there is no contradiction
between the results obtained with the low-resolution spectropolarimeter
FORS1 and those obtained with high-resolution spectropolarimeters. 
FORS1 is an instrument capable of performing reliable magnetic
field measurements, provided that the various source of 
uncertainties are properly taken into account.
}
\maketitle
%________________________________________________________________
\section{Introduction}
During a full decade of operations, the FORS1 instrument of the ESO
Very Large Telescope has collected a large number of magnetic field
measurements in various kinds of stars. Together with the ESPaDOnS
instrument of the Canada-France-Hawaii Telescope, and with the MuSiCoS and
NARVAL instruments of the 2\,m Telescope Bernard Lyot of the
Pic-du-Midi Observatory, FORS1 has been one of the workhorse
instruments for the observational studies of stellar magnetism.

Several important detections obtained with FORS1 have led to the
conclusion that magnetic fields are quite common in a variety of stars
across the Hertzsprung-Russell diagram, including for instance
central stars of planetary nebulae \citep{Joretal05},
hot subdwarfs \citep{Otoetal05}, 
$\beta$\,Cephei and slowly pulsating B stars \citep{Hubetal09a},
B stars with emission lines \citep{Hubetal09b},
and normal O-type stars \citep{Hubetal08b}.

In practice, a close inspection to the published results shows a number of problems:

\noindent
\textit{i)} Inconsistencies between field measurements obtained with
FORS1 and field measurements obtained with other instruments. For
instance, \citet{Hubetal04a} reported the discovery of a magnetic
field in the Herbig Ae/Be star HD\,139614, while repeated ESPaDOnS
measurements failed to confirm the magnetic nature of that star
\citep{Wadetal05}. Similarly, \citet{Siletal09} failed to confirm the
presence of a magnetic field in several of the $\beta$\,Cep and SPB stars
observed by \citet{Hubetal06a} and \citet{Hubetal09a}.

\noindent
\textit{ii)} Inconsistencies between the analysis of the same FORS
dataset performed by different authors. For instance, \citet{McSwain08}
observed the normal B stars NGC 3766 MG 111 and NGC 3766 MG 176, and 
the Be star NGC 3766 MG 200, and reported no field detection. 
Using the same data, \citet{Hubetal09b} reported new field
detections for all three stars.

\noindent
\textit{iii)} Inconsistencies between the analysis of the same FORS
dataset performed by the same authors but at different epochs. For
instance, \citet{Wadetal05} reported a possible detection in the young Ap
star HD\,72106A, which was not confirmed by the later analysis of the
same data performed by the same group \citep{Wadetal07a}. Note that the
magnetic nature of that star was established with data independently
obtained with ESPaDOnS \citep[see][]{Foletal08}.

\noindent
\textit{iv)} Inconsistencies between field measurements obtained from
different subsets of an observing series of frames. Magnetic field
measurements are often obtained by combining a number of pairs of
frames obtained at two different position angles of the retarder
waveplate. This redundancy is mainly motivated by the need to reach a
very high signal-to-noise ratio. In some rare cases, a magnetic field
may be firmly detected in a pair of frames, but not in the remaining
pairs.  This is for instance the case of the measurements of
HD\,139614 reported by \citet{Hubetal04a}, where the magnetic field is
detected only in a subset of frames, and in a couple of H Balmer lines.

\noindent
\textit{v)} Finally, there are some global inconsistencies of the full
FORS dataset, revealed for instance by the high incidence of field
detection in the null profiles; this kind of problem was not previously
reported in the literature, and will be discussed in
Sect.~\ref{Sect_Internal_Tests}.

Issues \textit{ii)} and \textit{iii)} point to possible glitches in
the data reduction method, while issues \textit{i)}, \textit{iv)}, and
\textit{v)} might be the symptom of a wider, possibly instrumental, problem.

The release of the FORS pipeline for spectropolarimetric data
\citep{Izzetal10,Izzetal11} gave us the opportunity to develop an accurate and
efficient reduction method. The FORS pipeline is a
software tool specifically designed for that instrument, which makes
it easier to handle some characteristics that are specific to the data
obtained with FORS1. Furthermore, the FORS pipeline allows
a high degree of automatization in the data reduction
process. Complemented with a suite of software tools for data
pre-processing (for frame classification and quality check) and data
post-processing (for the computation of the magnetic field), we were able to
build up a nearly automatic tool for data analysis which allowed us to
treat the entire FORS archive data in a homogeneous way. Using our
tool suite, the difference in terms of effort required to perform
the analysis of a single series of observations compared to performing the
analysis of the entire archive of FORS1 data consists mainly in the
amount of the necessary disk space and CPU time.

Compared to the reduction of raw data coming from individual observing
runs, the mass-production of reduced spectra offers the possibility to
perform a quality check of the final products on a very large scale.
The aim of this paper is to present: (1) a preliminary discussion of
the methods, (2) the results of our quality checks, and (3) to caution
the reader about the lack of robustness of some of the results previously
published in the literature. A deeper analysis and a comprehensive and
homogeneous catalogue of FORS1 magnetic field measurements will be
published in a forthcoming paper.

\section{Instrument and instrument settings}
FORS1 \citep{Appetal98} is a multi-purpose instrument capable of doing
imaging and low-resolution spectroscopy in the optical, equipped with
polarimetric optics. It was attached at the Cassegrain focus of
one of the 8\,m units of the ESO Very Large Telescope (VLT) of the
Paranal Observatory from the beginning of operations in 1999 until
instrument decommissioning in March 2009. The FORS1 polarimetric optics
were subsequently moved to the twin instrument FORS2.

Two detectors have been used in the FORS1 instrument: a 2k $\times$ 2k
Tektronix CCD (TK2048), (pixel size $24\,\mu$m $\times 24\,\mu$m),
and a mosaic composed of two 2k $\times$ 4k E2V CCDs (pixel size of
$15\,\mu$m$ \times 15\,\mu$m). TK2048 was used from the beginning of
operations to March 2007 (P65 to P78), E2V from April 2007 to the
decommissioning of the instrument in March 2009 (P79 to P82). For
more details about the two detectors, see \citet{Szeetal07}.

%%%%%%%%%%%%%%%%%%%%%%%%%%%%%%%%%%%%%%%%%%%%%%%%%%%%%%%%%%%%%%%%%%%%%%%%%%%%%%%%%%%%
\begin{table}
\caption{\label{Tab_Instrument_Settings} 
Summary of the characteristic of the grisms+CCD setting most commonly
employed for magnetic field measurements.}
\begin{tabular}{llcrc}
\hline \hline
Grism       & CCD    & Wavelength        & Dispersion     & Resolution \\
            &        & range (\AA)       & \AA\,pixel$^{-1}$& (1\arcsec slit width) \\
\hline
 600\,B     & TK2048 & 3480--5900        &    1.2           & 830  \\
 600\,B     & EEV    & 3400--6100        &    0.75          & 810 \\
1200\,g$^1$ & TK2048 & 4290--5470        &    0.58          & 1725 \\
1200\,B     & EEV    & 3660--5110        &    0.36          & 1540 \\
 600\,R$^2$ & TK2048 & 5250--7415        &    1.2           & 1230 \\
\hline
\end{tabular}
\smallskip

\noindent
1. We note that grism 1200\,g was often used setting the slit close to the right edge
of the instrument field of view. For that special setting, the observed wavelength
interval was $~3840 - 4970$\,\AA.\\
2. Grism 600\,R was used together with order separation filter GG\,435.
\end{table}
%%%%%%%%%%%%%%%%%%%%%%%%%%%%%%%%%%%%%%%%%%%%%%%%%%%%%%%%%%%%%%%%%%%%%%%%%%%%%%%%%%%%
Most of the observing programmes used the so-called ``fast-mode''.
In this mode the target is just one star in the centre of the instrument field of
view.  A small number of observing programmes used the multi-object
mode, which allows the simultaneous recording of the polarized spectra
for up to nine different stars within a $\sim 7\arcmin \times 7\arcmin$
field of view \citep[see e.g.][]{Bagetal06}.  In multi-object mode,
the FORS pipeline is not yet robust enough to correctly associate the
parallel and the perpendicular beams of the same target, especially
when the observations are performed with the grism 600\,B. The reason
is due to the presence of a number of ghost images (probably coming
from the Longitudinal Atmospheric Dispersion Corrector), which are
interpreted by the pipeline as stellar spectra. The solution to this
problem would be to force the pipeline to extract only user selected
spectra, an option that has not been implemented yet. Therefore we did
not re-reduce the large majority of the observations performed in
multi-object mode.

Most of the observations were performed with grism 600\,B, some with
grism 1200\,B, and a small number with grism 1200\,g and 600\,R (see
Table~\ref{Tab_Instrument_Settings}).  In most of the cases, the slit
width was set to 0.4\arcsec\ or 0.5\arcsec, for a typical spectral
resolution of about 1600 -- 2300 (with grism 600\,B or 600\,R) or 3000
-- 3400 (with grism 1200\,g or 1200B).

\section{Data reduction}
In the following, we will adopt the same formalism used in
\citet{Bagetal09}; \fo\ and \fe\ are the fluxes in the parallel
and in the perpendicular beam of the polarisation analyser,
respectively, $\pv\ = V/I$ is the circular polarization normalised to
the intensity, and \nv\ is the null profile, a quantity that was
introduced by \citet{Donetal97}, and that is representative of the
noise of \pv.

We always obtained the \pv\ profiles from a series of one or more
pairs of exposures. Each pair of exposures is composed of two frames
obtained with the retarder waveplate at position angles separated by
90\degr. For most of the observing series, it was also possible to
calculate the null profile.  For those cases in which the number of
pairs of exposures $N$ was odd and $\ge 3$, the null profile was
obtained omitting the last pair of exposures. Obviously, with
just one pair of exposure, the null profile could not be calculated.

The extracted fluxes were combined to obtain the \pv\ and \nv\ profiles 
using the formulas of the difference methods given in Eqs.~(A2) and (A7) 
of \citet{Bagetal09}, which for convenience we reproduce below:
%%%%%%%%%%%%%%%%%%%%%%%%%%%%%%%%%%%%%%%%%%%%%%%%%%%%%%%%%%%%%%%
\begin{equation}
\begin{array}{rcl}
\pv &=& {1 \over 2 N} \sum\limits_{j=1}^N \left[ 
\left(\frac{\fo - \fe}{\fo + \fe}\right)_{\alpha_j} - 
\left(\frac{\fo - \fe}{\fo + \fe}\right)_{\alpha_j + \dalphao}\right] \\[2mm]
\nv &=& {1 \over 2 N} \sum\limits_{j=1}^N (-1)^{(j-1)}\left[ 
\left(\frac{\fo - \fe}{\fo + \fe}\right)_{\alpha_j} - 
\left(\frac{\fo - \fe}{\fo + \fe}\right)_{\alpha_j + \dalphao}\right]\; ,\\
\end{array}
\label{Eq_V_and_N}
\end{equation}
%%%%%%%%%%%%%%%%%%%%%%%%%%%%%%%%%%%%%%%%%%%%%%%%%%%%%%%%%%%%%%%
where $\dalphao=90^\circ$ and $\alpha_j$ belongs to the set $
\{-45^\circ$, $135^\circ \}$.\footnote{We note that instead of setting
  the $\lambda/4$ retarder waveplate to all four possible angles, most
  of the observers preferred to use only the angles $-45^\circ$ and
  $+45^\circ$.}  However, in this work we adopt an important
modification as described below.

\subsection{Rectifying Stokes profiles}
In some cases we found the \pv\ profile clearly offset from zero, even
when no circular polarization of the continuum is expected.  This
happened for instance in Herbig Ae/Be stars observed by
\citet{Wadetal07a}, and in several other cases in the course of the
present work.

A possible explanation is cross-talk from linear to circular
polarization, as discussed by \citet{Bagetal09}. Cross-talk may be a
problem only if the observed source is linearly polarised, and
is especially significant for spectra acquired with a slitlet close to the
edge of the instrument field of view (as in some observations
obtained in multi-object mode). However, we often found slight but
noticeable offsets also in FORS data for stars that are not linearly
polarised, and that were observed in the centre of the field of view. These
offsets may be explained if the ratio
between the transmission functions in the perpendicular beam
$h^\perp$, and the transmission function in the parallel beam,
$h^\parallel$, is not constant as the retarder waveplate is set to
the different position angles.

Inspection of the null profile may help to discriminate between the two
cases. If \pv\ is offset from zero due to cross-talk from linear
polarization (or in fact because the source is intrinsically
circularly polarized in the continuum), the null profile will still be
oscillating about zero, while an offset introduced by a non constant
ratio $h^\perp/h^\parallel$ will still appear (somewhat scaled down)
in the null profile. This can be seen analytically as follows.  We
first recall that within the framework of the difference method, the
ratio $h = h^\perp/h^\parallel$ is developed to first order:
%%%%%%%%%%%%%%%%%%%%%%%%%%%%%%%%%%%%%%%%%%%%%%%%%%%%%%%%%%%%%%%
\begin{equation}
h = \frac{h^\perp}{h^\parallel} = 1 + \dhh\ .
\label{Eq_dhh}
\end{equation}
%%%%%%%%%%%%%%%%%%%%%%%%%%%%%%%%%%%%%%%%%%%%%%%%%%%%%%%%%%%%%%%
We then assume that the term \dhh\ that appears in Eqs.~(34) of
\citet{Bagetal09} depends on the position angle \alphao\ of the
retarder waveplate, and we repeat the computations that lead to
their Eqs.~(37). We obtain
%%%%%%%%%%%%%%%%%%%%%%%%%%%%%%%%%%%%%%%%%%%%%%%%%%%%%%%%%%%%%%%
\begin{equation}
\begin{array}{rcl}
\pvr &=& {1 \over 2 N} \sum\limits_{j=1}^N \left[ 
\left(\frac{\fo - \fe}{\fo + \fe}\right)_{\alpha_j} - 
\left(\frac{\fo - \fe}{\fo + \fe}\right)_{\alpha_j + \dalphao} - a_j\right] b_j^{-1} \\[2mm]
\nvr &=& {1 \over 2 N} \sum\limits_{j=1}^N (-1)^{(j-1)}\left[ 
\left(\frac{\fo - \fe}{\fo + \fe}\right)_{\alpha_j} - 
\left(\frac{\fo - \fe}{\fo + \fe}\right)_{\alpha_j + \dalphao} - a_j\right] b_j^{-1}\; ,\\
\end{array}
\label{Eq_Rectified}
\end{equation}
%%%%%%%%%%%%%%%%%%%%%%%%%%%%%%%%%%%%%%%%%%%%%%%%%%%%%%%%%%%%%%%
where
%%%%%%%%%%%%%%%%%%%%%%%%%%%%%%%%%%%%%%%%%%%%%%%%%%%%%%%%%%%%%%%
\begin{equation}
\begin{array}{rcl}
a_j &=&  \frac{2 \left[\dhh(\alpha_j + \dalphao) - \dhh(\alpha_j)\right]}
              {4 + 2 \left[\dhh(\alpha_j) + \dhh(\alpha_j + \dalphao)\right] + 
                  \dhh(\alpha_j)\, \dhh(\alpha_j + \dalphao)}\\[1mm]
b_j &=&  \frac{4 + 2 \left[\dhh(\alpha_j) + \dhh(\alpha_j + \dalphao)\right]}
              {4 + 2 \left[\dhh(\alpha_j) + \dhh(\alpha_j + \dalphao)\right] + 
                  \dhh(\alpha_j)\, \dhh(\alpha_j + \dalphao)}\\
\end{array}
\label{Eq_ak_bk}
\end{equation}
%%%%%%%%%%%%%%%%%%%%%%%%%%%%%%%%%%%%%%%%%%%%%%%%%%%%%%%%%%%%%%%
and where we have introduced the new symbols \pvr\ and \nvr\ to denote
the \pv\ and \nv\ profiles rectified.

Equation~(\ref{Eq_Rectified}) shows that if $\dhh(\alpha_j)\,
\dhh(\alpha_j + \dalphao) \ll 1$, then $b_j \simeq 1$, and \pvr\ and
\nvr\ can be simply obtained by subtracting from the \pv\ and
\nv\ profiles obtained from Eqs.~(\ref{Eq_V_and_N}) two smooth curves
that interpolate the continuum of those profiles. These smooth curves
may be obtained fitting a polynomial to \pv\ and to \nv, or using a
Fourier filter. Following this simple method, \pv\ and \nv\ are
rectified independently, and \nv\ loses part of its diagnostic
content, because it is forced to oscillate about zero in a
way that is totally independent of \pv. An alternative method consists of
estimating the transmission functions \dhh\ by smoothing the ratios
between fluxes on the parallel and perpendicular beams, then obtaining
the explicit expressions for the individual $a_j$ and $b_j$
coefficient, and finally using Eqs.~(\ref{Eq_Rectified}) to compute the
\pvr\ and \nvr\ profiles in a consistent way. In practice, we found that
the two alternative methods lead always to very similar results.

\subsection{Clipping with the null profile}\label{Sect_Clipping}
Both \pv\ and \nv\ profiles show occasional spikes that occur in 
the same wavelength bin. Most of these spikes are probably
produced by cosmic ray hits, and, if not removed, may lead to
spurious detections of magnetic fields, or at least decrease the
precision of its determination. A possible remedy is to exploit the
null profiles to clip the \pv\ profiles by discarding from the
computation of the \bz\ values those points for which the (rectified)
\nv\ value depart from zero by more than $k\,\sigma$, where $k$ is a
constant typically $\sim 3$.  The clipping may be applied to the individual
deviating points, or may also remove the adjacent points.

\subsection{Magnetic fields determinations}\label{Sect_Field_Determinations}
%%%%%%%%%%%%%%%%%%%%%%%%%%%%%%%%%%%%%%%%%%%%%%%%%%%%%%%%%%%%%%%%%%%%%%%%%%%
\begin{figure*}
\rotatebox{270}{\scalebox{0.68}{
\includegraphics*[0.8cm,0.0cm][6cm,28cm]{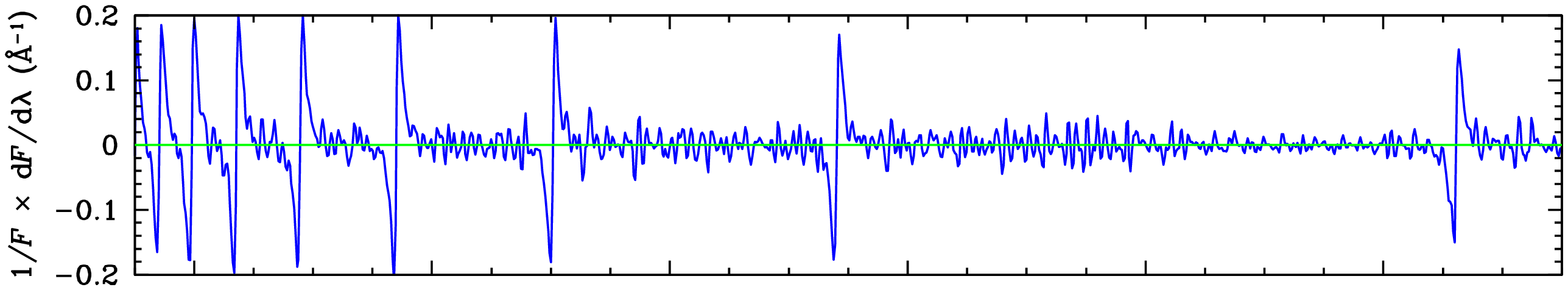}\\
\includegraphics*[1.0cm,0.0cm][21cm,28cm]{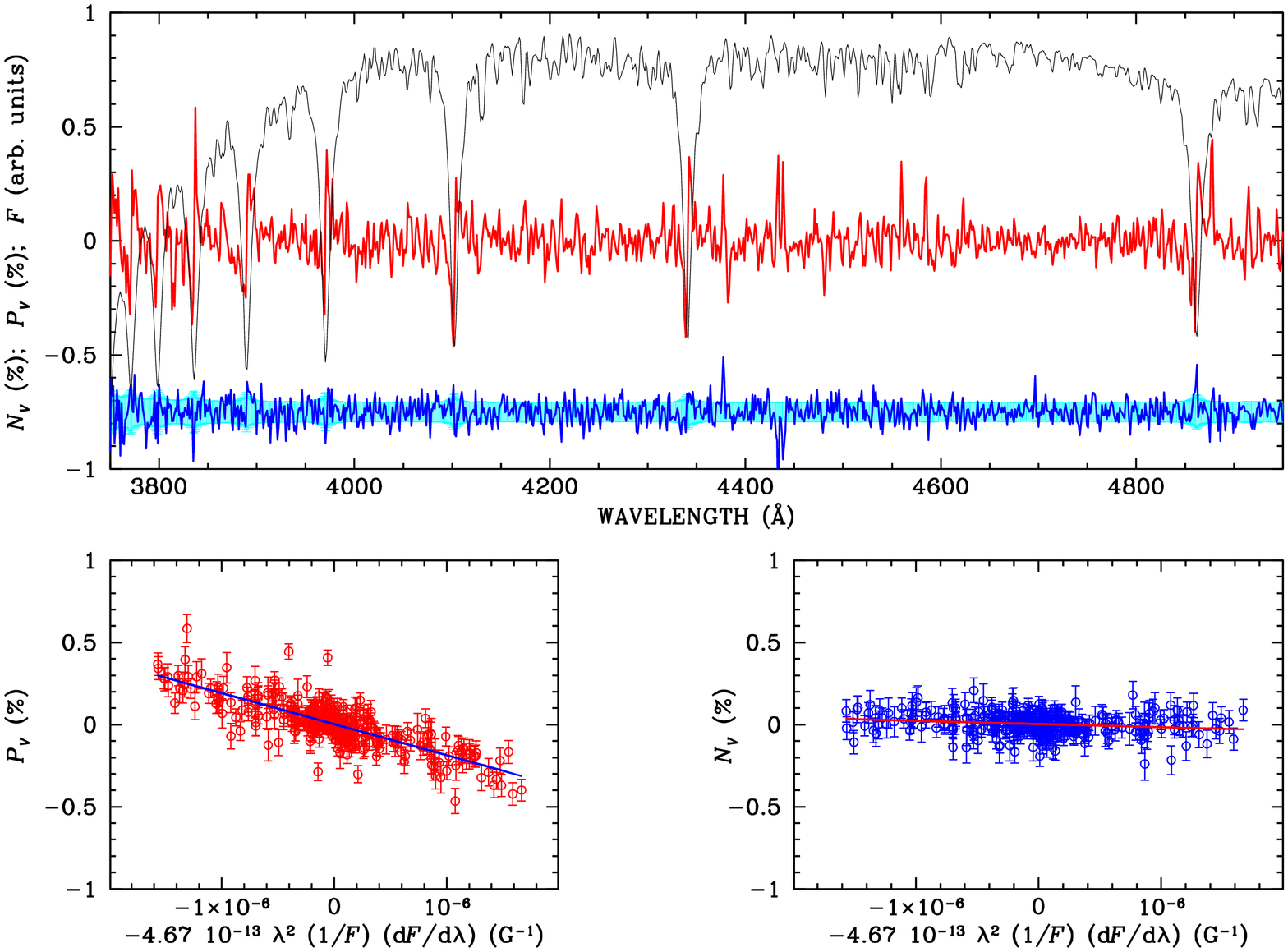}}}
\caption{\label{Fig_HD94660} 
The polarised spectrum of HD\,94660 observed on 2007-03-22 with
grism 600B.
The top panel shows the normalised derivative of the observed flux, and
the panel immediately below
shows the observed flux $F$ (black solid line, in arbitrary units,
and not corrected for the instrument response), the \pv\ profile
(red solid line
centred about 0), and the null profile (blue solid line, offset by
--0.75\,\% for display purpose). The null profile is
expected to be centred about zero and scattered according to a gaussian
with $\sigma$ given by the \pv\ error bars. The \pv\ error bars
are represented with light blue bars centred about --0.75\,\%.
The slope of the interpolating lines in the bottom
panels gives the mean longitudinal field from \pv\ (left bottom panel) and
from the null profile (right bottom panel) both calculated using the H
Balmer lines only. The corresponding \bz\ and \nz\ values
are $-1885 \pm 90$\,G and $-192 \pm 62$\,G, respectively.
}
\end{figure*}
%%%%%%%%%%%%%%%%%%%%%%%%%%%%%%%%%%%%%%%%%%%%%%%%%%%%%%%%%%%%%%%%%%%%

FORS longitudinal magnetic field measurements are obtained using the relationship
%%%%%%%%%%%%%%%%
\begin{equation}
V(\lambda)= - g_\mathrm{eff} \ \cz \ \lambda^{2} \
%                \frac{1}{I(\lambda)} \
                \frac{\mathrm{d}I(\lambda)}{\mathrm{d}\lambda} \
                \bz\;,
\label{Eq_Bz}
\end{equation}
%%%%%%%%%%%%%%%%
where $V(\lambda)$ and $I(\lambda)$ are the Stokes $V$ and $I$
profiles of a spectral line, $g_\mathrm{eff}$ is the effective Land\'{e} factor, and
%%%%%%%%%%%%%%%%
\begin{equation}
\cz = \frac{e}{4 \pi m_\mathrm{e} c^2}
\ \ \ \ \ (\simeq 4.67 \times 10^{-13}\,\AA^{-1}\ \mathrm{G}^{-1})
\end{equation}
%%%%%%%%%%%%%%%%
where $e$ is the electron charge,
$m_\mathrm{e}$ the electron mass, and $c$ the speed of light.

\citet{Bagetal02} proposed to use a least-squares technique to derive
the longitudinal field via Eq.~(\ref{Eq_Bz}), by minimising the
expression
%%%%%%%%%%%%%%%%
\begin{equation}
\chi^2 = \sum_i \frac{(y_i - \bz\,x_i - b)^2}{\sigma^2_i}
\label{Eq_ChiSquare}
\end{equation}
%%%%%%%%%%%%%%%%
where, for each spectral point $i$,
$y_i = \pv(\lambda_i)$,
$x_i = -g_\mathrm{eff}\ \cz \lambda^2_i (1/F\ \times \mathrm{d}F/\mathrm{d}\lambda)_i$,
$F_i$ is the flux measured in the spectral bin at $\lambda_i$, and 
$b$ is a constant introduced to
account for possible spurious residual polarization in the
continuum. Note that the spurious polarization is assumed constant in
wavelength, which may not be true if we adopt \pv\ from 
Eqs.~(\ref{Eq_V_and_N}), but it is a realistic assumption if we use
\pvr\ of Eq.~(\ref{Eq_Rectified}), for which we must retrieve $b = 0$
within its error bar.

Equation~(\ref{Eq_Bz}) is subject to several important limitations.
(i) It is valid only in the limit of fields weak enough
that Zeeman splitting is small compared to the {\em local} spectral
line width, typically with a $\la 1$\,kG strength in the case of
optical metal spectral lines, or $\la 10$\,kG for Balmer
lines. (ii) It applies to isolated, unblended lines.  (iii)
The value of $g_{\rm eff}$ varies by a significant amount from line to
line; the use of an average value means that for many lines, the actual
circular polarization will vary from the average one by up to
$~25$\,\%. The impact of these limitations was discussed
by \citet{Bagetal02}; they are not important if one is simply interested 
in field detection, or to asses the magnitude of the longitudinanal
field, but they should be kept in mind whenever FORS measurements are
used for modelling purposes. 

When we derive \bz\ by minimizing the $\chi^2$ of Eq.~(\ref{Eq_ChiSquare}),
we make the implicit assumption
%%%%%%%%%%%%%%%%%%%%%%%%%%%%%%%%%%%%%%%%%%%%%%%%%%%%%%%%%%%%%%%%
\begin{equation}
\frac{1}{I(\lambda)} \ \frac{\mathrm{d}I(\lambda)}{\mathrm{d}\lambda} =
\left(\frac{1}{F}        \ \frac{\mathrm{d}F}{\mathrm{d}\lambda}\right)_i  \ ,
\end{equation}
%%%%%%%%%%%%%%%%%%%%%%%%%%%%%%%%%%%%%%%%%%%%%%%%%%%%%%%%%%%%%%%%
which is justified provided that
%%%%%%%%%%%%%%%%%%%%%%%%%%%%%%%%%%%%%%%%%%%%%%%%%%%%%%%%%%%%%%%%
\begin{equation}
\vert \frac{1}{I(\lambda)} \ \frac{\mathrm{d}I(\lambda)}{\mathrm{d}\lambda} \vert \gg
\vert \frac{1}{T(\lambda)} \ \frac{\mathrm{d}T(\lambda)}{\mathrm{d}\lambda} \vert
\end{equation}
%%%%%%%%%%%%%%%%%%%%%%%%%%%%%%%%%%%%%%%%%%%%%%%%%%%%%%%%%%%%%%%%
%%%%%%%%%%%%%%%%%%%%%%%%%%%%%%%%%%%%%%%%%%%%%%%%%%%%%%%%%%%%%%%%%%%%%%%%%%%
\begin{figure*}
\rotatebox{270}{\scalebox{0.68}{
\includegraphics*[1.0cm,0.0cm][20.5cm,28cm]{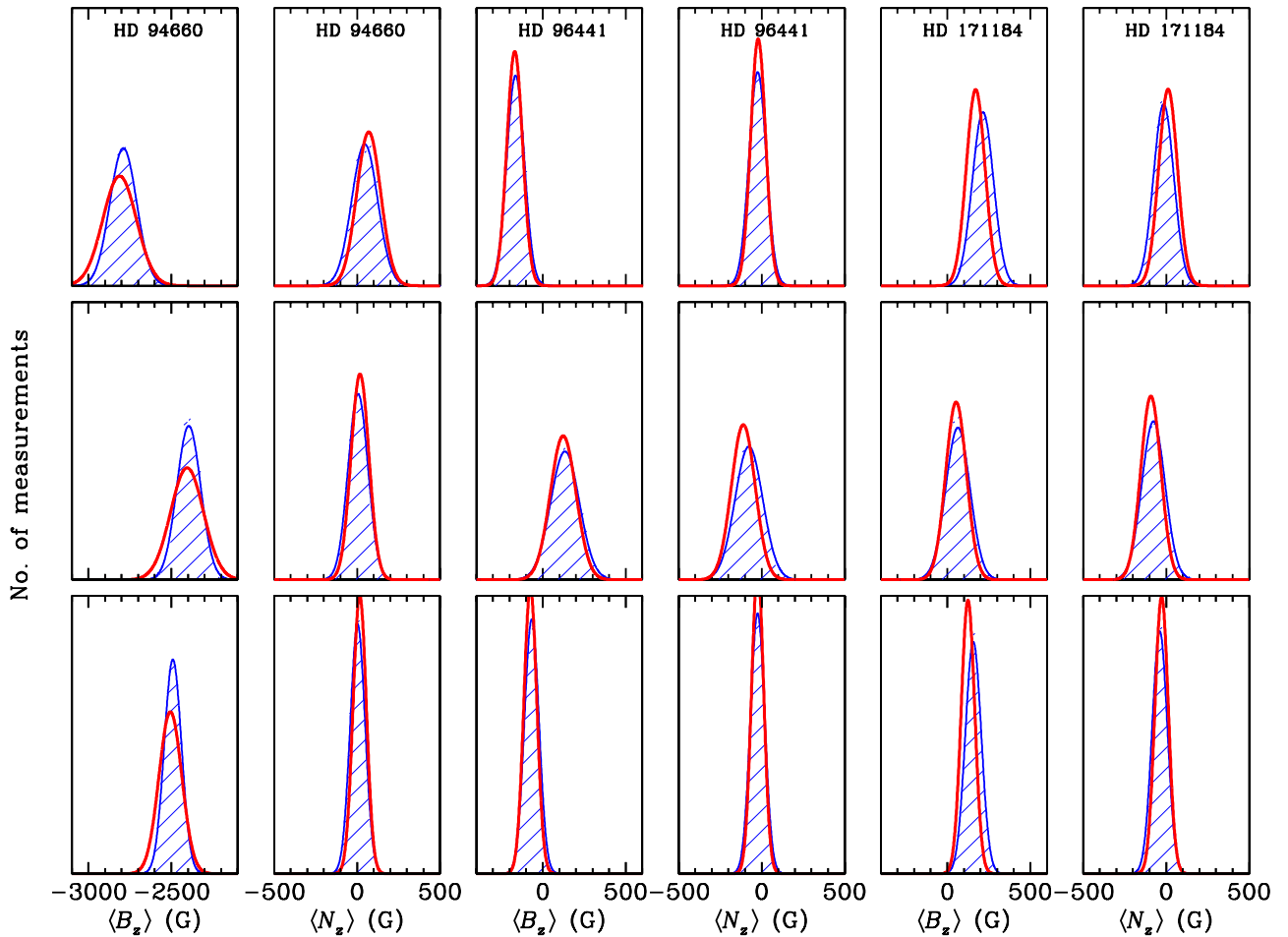}}}
\caption{\label{Fig_Montecarlo} 
Results obtained from the analysis of
  the H Balmer lines (top panels), of the metal lines (mid panels), and
  of the full spectrum (bottom panels) for
  three stars, using standard error theory and Montecarlo
  simulations. The (red) thick solid lines are Gaussian curves centred about
the various field values and with $\sigma$ given by Eq.~(\ref{Eq_External_Error}).
The blue lines show the distributions obtained with Monte Carlo simulations
as explained in the text. Note that HD~94660 has a large field, HD~96441 has no detected field, and HD~171184 has
a weak but probably real field. 
}
\end{figure*}
%%%%%%%%%%%%%%%%%%%%%%%%%%%%%%%%%%%%%%%%%%%%%%%%%%%%%%%%%%%%%%%%%%%%
where $T(\lambda)$ is a function that accounts for the stellar
continuum, the transmission functions of the interstellar medium, of
the Earth atmosphere, of the telescope, and of the instrument. The
validity of this approximation can be empirically evaluated case by
case by verifying that the impact of the continuum rectification to
the normalised derivative of Stokes $I$ is negligible. Practically, a
quick check can be performed even without continuum normalization.
Since the function $T(\lambda)$ is much smoother
than $I(\lambda)$, the contributions due to the terms
$[1/I(\lambda)\ \times\ \mathrm{d}I(\lambda)/\mathrm{d}\lambda]$ and
$[1/T(\lambda)\ \times\ \mathrm{d}T(\lambda)/\mathrm{d}\lambda]$ to the
quantity
$[1/F\ \times\ \mathrm{d}F/\mathrm{d}\lambda]_i$
can be visually disentangled simply by plotting the function
$[1/F\ \times\ \mathrm{d}F/\mathrm{d}\lambda]_i$ versus $\lambda$
over a zero line.
The top panel of Fig.~\ref{Fig_HD94660} shows
that the broad contribution due to $T(\lambda)$ 
is totally negligible compared to the fine
structure due to $I(\lambda)$, and that 
$[1/F\ \times\ \mathrm{d}F/\mathrm{d}\lambda]_i$ practically
oscillates about zero. For hot stars observed with grism 1200\,B,
$T(\lambda)$ is very steep, and 
$[1/T(\lambda)\ \times\ \mathrm{d}T(\lambda)/\mathrm{d}\lambda]$
may depart from zero more significantly than what displayed in
Fig.~\ref{Fig_HD94660}, yet the impact of the continuum normalisation
on the \bz\ determination is still within the error bars due to
photon-noise.

Finally, we note that for quality check purpose, it is useful to
measure the mean longitudinal field by using the null
profiles instead of the \pv\ profiles. This \textit{null field} value
\nz\ should be found consistent with zero within its error
bars. 

Figure~\ref{Fig_HD94660} illustrates the technique discussed in
this Section, showing the results obtained for the FORS
observations of the well known magnetic star HD\,94660.

\subsection{Error bars} \label{Sec_Error_Bars}
\citet{Bagetal02} have obtained the error bar of
the longitudinal magnetic field from the inverse of the $\chi^2$
matrix. This way, the ``internal'' error bar $s_{\bz}$ was obtained
by propagating the errors $\sigma_i$ of the \pv\ profiles onto the analytical
expression of \bz\ that is obtained by minimising the $\chi^2$
expression of Eq.~(\ref{Eq_ChiSquare}):
%%%%%%%%%%%%%%%%%%%%%%%%%%%%%%%%%%%%%%%%%%%%%%%%%%%%%%%%%%%%%%%%%
\begin{equation}
s^2_{\bz} = \frac{ \left(\sum_i \frac{1}{\sigma^2_i}\right) } 
        {  \left(\sum_i \frac{1}{\sigma^2_i}\right) 
           \left(\sum_i \frac{x_i^2}{\sigma^2_i}\right)
                - \left(\sum_i \frac{x_i}{\sigma^2_i}\right)^2 }\ ,
\label{Eq_Internal_Error}
\end{equation}
%%%%%%%%%%%%%%%%%%%%%%%%%%%%%%%%%%%%%%%%%%%%%%%%%%%%%%%%%%%%%%%%%
where $\sum_i$ is a sum extended to all $i$ spectral bins used
in Eq.~(\ref{Eq_ChiSquare}). An analogous expression holds for
the error on the null field \nz, $s_{\nz}$, which turns to be
numerically \textit{identical} to $s_{\bz}$, since $\sigma^2_i$ 
are identical for the \pv\ and \nv\ profiles. 

Since there are numerous indications that the error bars of FORS1
field measurements are underestimated, in this work we introduce an
``external'' error bar \sbz\ that takes into account the actual scattering
about the interpolating line, i.e.:
%%%%%%%%%%%%%%%%%%%%%%%%%%%%%%%%%%%
\begin{equation}
\sbz = s_{\bz} \sqrt{\chi^2_{\rm min}/\nu} 
\label{Eq_External_Error}
\end{equation}
%%%%%%%%%%%%%%%%%%%%%%%%%%%%%%%%%%%
where $\chi^2_{\rm min}$ is the minimum value of the $\chi^2$ of
Eq.~(\ref{Eq_ChiSquare}) and $\nu$ is the number of degrees of freedom
of the system, i.e., the number of spectral points minus two.
Strictly speaking, the use of Eq.~(\ref{Eq_External_Error}) is not
theoretically justified. It is equivalent to the hypothesis that
Eq.~(\ref{Eq_Bz}) is correct, but that the errors $\sigma_i$ on
\pv\ are all underestimated by a constant value $(\chi^2_{\rm
  min}/\nu)^{1/2}$.

The reality is that, as discussed above, Eq.~(\ref{Eq_Bz}) is not
rigorously valid, especially for strong fields. Departures from the
behaviour predicted by Eq.~(\ref{Eq_Bz}) mean that points in the
regression (as shown in the lower panels of Fig.~\ref{Fig_HD94660})
will deviate from the interpolating line by more than the amount
predicted by photon statistics, and the reduced $\chi^2$ values will
be $> 1$. For stars with weak or no magnetic field, or if the field is
computed from the null profiles, the scatter in the positions of
individual points around the regression line is dominated by noise
rather than by the approximations implicit in Eq.~(\ref{Eq_Bz}). In
this case the reduced $\chi^2$ is expected to be very close to 1, and
the ``internal'' and ``external'' errors tend to be quite similar. The
advantage of systematically using Eq.~(\ref{Eq_External_Error}) is to
automatically take into account a few cases where even if the field is
weak or non-existent, the reduced $\chi^2$ is $>1$.

To check the validity of our error estimates we have performed some
Monte Carlo simulations.  All pixel values of the raw frames of an
observing series were scattered, using a random number generator,
according to a Gaussian distribution with $\sigma$ given by the square
root of their electron counts.  This altered dataset was reduced, and
an estimate of the \bz\ and \nz\ values was obtained. The same
procedure was repeated 5000 times to obtain a distribution for both
\bz\ and \nz\ values. Then, the centres of these distributions
$\bz^{MC}$ and $\nz^{MC}$, and their standard deviations $\sbz^{\rm MC}$
and $\snz^{\rm MC}$ were compared to the \bz, \nz, \sbz, \snz, $s_{\bz}$
(= $s_{\nz}$) values obtained from the observing series not altered by
the Monte Carlo simulation. Results for three observing series are
shown in Fig.~\ref{Fig_Montecarlo}. For the (large) \bz\ values of
HD\,94660 we found $\sbz > s_{\bz}$ and $\sbz > \sbz^{\rm
  MC}$; for all the remaining cases the error bars coming from the
three methods are similar.

\subsection{Additional details about data reduction}\label{Sect_Details}
In this Section we briefly discuss some issues related to data
reduction which influence the estimates of the mean longitudinal
magnetic field and their uncertainties, i.e., \textit{i)} flat-field
correction; \textit{ii)} the algorithm for spectrum extraction;
\textit{ii)} spectral rebinning; \textit{iv)} rectification of the
\pv\ and \nv\ profiles; and \textit{v)} the choice of the Land\'e
factor and of the spectral region used to determine the magnetic
field.

This (non-exhaustive) list gives an idea of the
multiple choices that are either implicitly or expliciting taken
during the process of data reduction. The following discussion helps to
understand how much the final results may depend on the actual
algorithm adopted for data reduction.

\noindent
\textit{i)} While flat-fielding is not needed to measure Stokes
profiles \citep[see, e.g.,][]{Bagetal09}, pixel-to-pixel variation may
have an impact on the evaluation of the $x_i$ term of
Eq.~(\ref{Eq_ChiSquare}), as pointed out by
\citet{Rivetal10}. However, in most of the cases, the signal-to-noise
ratio of the combined flat field frames is not much higher than that
of the science frames.  Furthermore, calibration frames show internal
reflections (especially in the bluest regions), but with different
patterns than seen in science frames. In some cases, it is clear that
internal reflections observed in the flat-field affect the
\bz\ determination, e.g., the difference between cases (a) and (b) of
Table~\ref{Tab_Reductions} described later in this Section is entirely
due to spurious features of the master screen flat. In
summary, it is not possible to conclude that flat-fielding correction
improves the longitudinal field determination, at least with the CCDs
that were used in FORS1.  We found however that the difference between
magnetic field values obtained with and without flat-fielding
corrections were generally well within photon-noise error bars.
%%%%%%%%%%%%%%%%%%%%%%%%%%%%%%%%%%%%%%%%%%%%%%%%%%%%%%%%%%%%%%%%%%%%%%%%%%%
\begin{figure}
\scalebox{0.4}{
\includegraphics*[0cm,5cm][22cm,25.5cm]{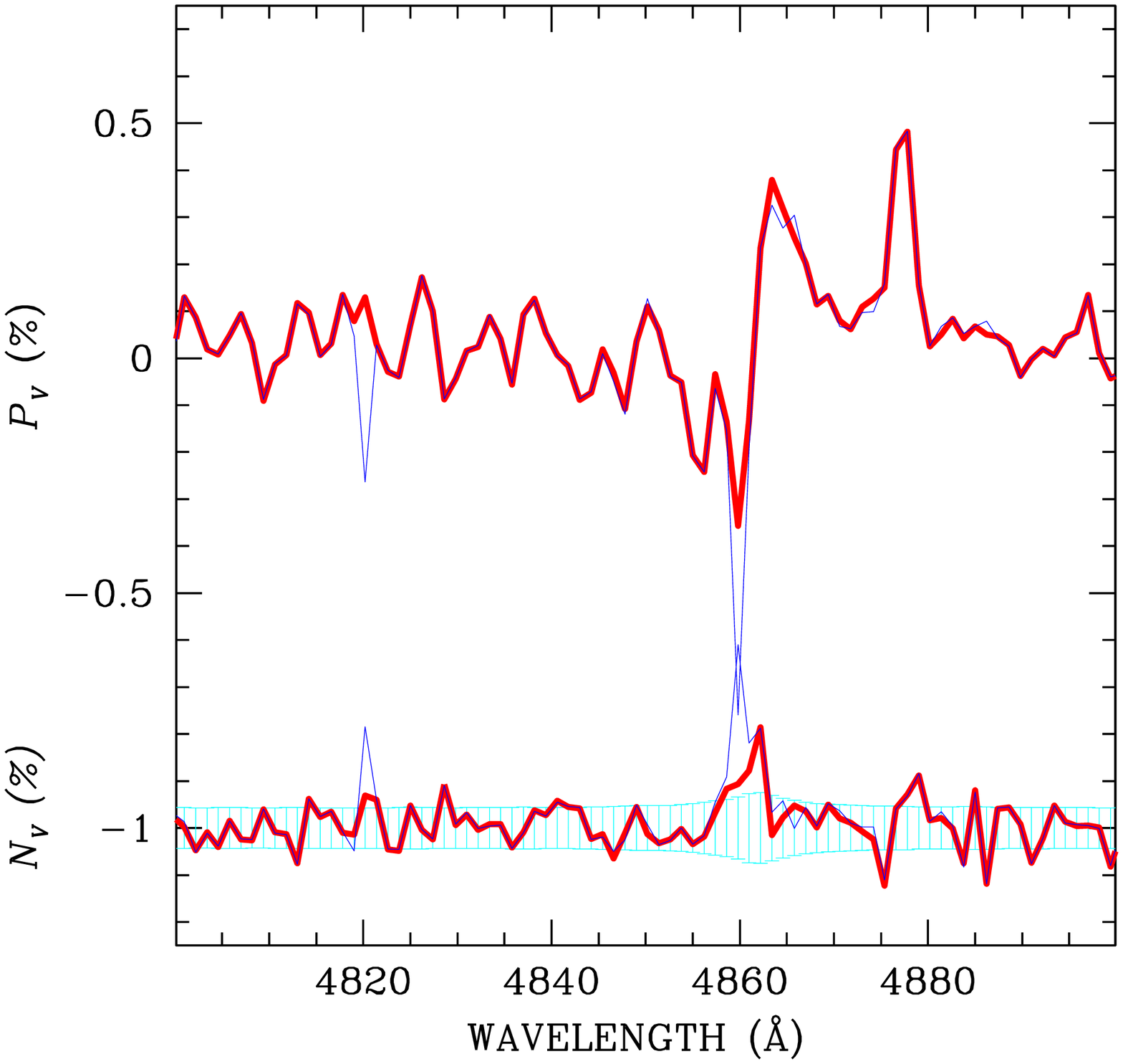}}
\caption{\label{Fig_Optimal_vs_Aperture} \pv\ and \nv\ profiles of
HD\,94660 obtained with average extraction (red thick lines) an 
optimal extraction (blue thin lines). The null profiles are offset
by -1\,\% for display purpose, and overplotted to the statistical error
bars of the \pv\ and \nv\ profiles.}
\end{figure}
%%%%%%%%%%%%%%%%%%%%%%%%%%%%%%%%%%%%%%%%%%%%%%%%%%%%%%%%%%%%%%%%%%%%
%%%%%%%%%%%%%%%%%%%%%%%%%%%%%%%%%%%%%%%%%%%%%%%%%%%%%%%%%%%%%%%%%%%%%%%%%%%
\begin{figure}
\scalebox{0.5}{
\includegraphics*[1.0cm,5.5cm][22cm,25.5cm]{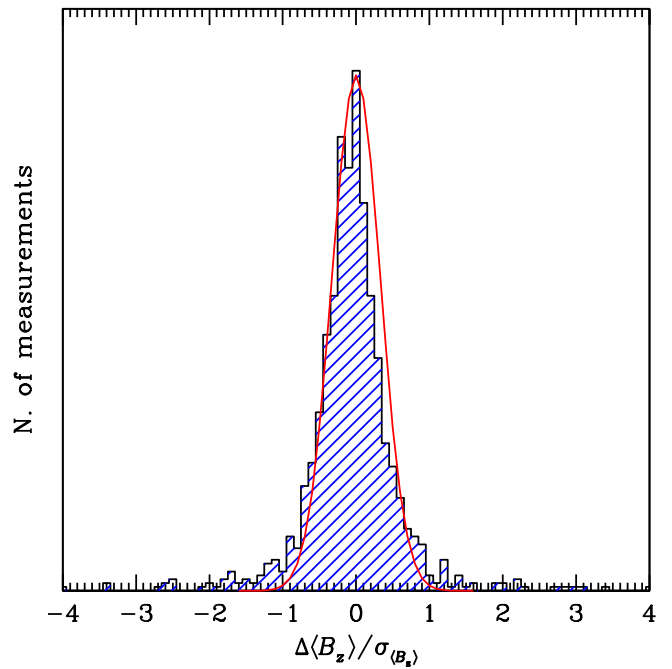}}
\caption{\label{Fig_Histo_Reductions} The distribution of the difference between 
\bz\ values obtained with two slightly different algorithms for data reduction,
normalised to the error bars, and compared to a Gaussian with identical area,
centred about zero, and with $\sigma=0.33$.
}
\end{figure}
%%%%%%%%%%%%%%%%%%%%%%%%%%%%%%%%%%%%%%%%%%%%%%%%%%%%%%%%%%%%%%%%%%%%

\noindent
\textit{ii)} The FORS pipeline \citep{Izzetal11} offers two different
options for spectral extraction: optimal extraction \citep{Horne86},
and aperture extraction. Using the former method, each pixel in the
spectra extraction is weighted according to the fraction of the flux
which is expected assuming a model of the spatial profile across the
dispersion. The method substantially increases the signal-to-noise
ratio when the noise is dominated by background, and automatically
accounts for cosmic-rays, but may be less suitable than ``average
extraction'' when the noise is determined mainly by Poisson
statistics. We found that in general, the polarized spectra obtained
with optimal extraction have numerous spurious spikes; an example is
shown in Figure~\ref{Fig_Optimal_vs_Aperture} with the detail of the
polarized spectrum of HD\,94660 reduced with average extraction (blue
thick lines) and with optimal extraction (red thin lines). This
suggests that the aperture extraction should be preferred to optimal
extraction. Note that FORS pipeline assumes a certain default aperture
which may not be the optimal one for all cases. For our data 
re-reduction we generally adopted a 12 pixel aperture, using
a larger one for a few cases of very bad seeing.

\textit{iii)} Generally, the wavelength bin size of the calibrated
spectra is set equal to the dispersion (measured as \AA\ per pixel)
pertaining to each given instrument setting, but there is no
obvious reason why one should not experiment with an interpolation
over a more refined wavelength grid, or with a rebinning
to a larger bin size to increase the signal-to-noise ratio of each
point. Surprisingly, the final \bz\ value depends on the
adopted choice of the bin size, in a way that deserves further
investigation.

\textit{iv)} The choice whether to use Balmer lines only or the full
spectrum, and the choice for the Land\'e factor that best represents
the average Zeeman pattern of the spectral lines may be different from
work to work, and therefore lead to different numerical values for
\bz. However, since these choices affect both the field value and its error bar
in the same way, they will not change a detection into a non-detection
nor vice-versa. For more details on how field results change when
Eq.~(\ref{Eq_ChiSquare}) is applied to Balmer lines only, or only to
He+metal lines, or to the full spectrum, see, e.g., \citet{Bagetal02}
and \citet{Bagetal06}.

\textit{v)} In some cases, using the rectified profiles of
Eq.~(\ref{Eq_Rectified}) instead of the non-rectified \pv\ profiles of
Eq.~(\ref{Eq_V_and_N}) leads to significant changes in the field values,
and may even turn a field detection into a non-detection, but in most
of the cases, field changes are within the error bars.

Table~\ref{Tab_Reductions} shows the field values calculated for three
stars using various algorithms. It is interesting to note that a field
measurement of the Ap star HD\,171184 could be found to be either as a
non detection, a marginal detection, or even a $4\,\sigma$ detection,
according to the selected reduction algorithm (since the star is
chemically peculiar, probably the star has a weak magnetic field). As
a further example of the impact of a choice made during data
reduction, Fig.~\ref{Fig_Histo_Reductions} shows the results the
distribution of the difference between \bz\ values obtained with
methods (a) and (e) of Table~\ref{Tab_Reductions}, normalised to the
error bars, and compared with a Gaussian with the same area and
$\sigma=0.33$.

While it is certainly possible to argue that certain algorithms for
data reduction lead to more robust results than other methods, some
alternative choices may be equally reasonable. The fact that there
are alternative and reasonable choices during the process of data
reduction adds a contribution to the error bar of the mean
longitudinal field. This contribution is difficult to quantify in
general, but it is a non-negligible fraction of the error bar from photon noise.

%%%%%%%%%%%%%%%%%%%%%%%%%%%%%%%%
\begin{table}
\caption{\label{Tab_Reductions} Longitudinal magnetic fields of three
stars obtained with different algorithms.}
\begin{center}
\begin{tabular}{lr@{$\pm$\,}lr@{$\pm$\,}lr@{$\pm$\,}l}
\hline\hline
                             &
\multicolumn{2}{c}{HD 94660} &
\multicolumn{2}{c}{HD 96441} &
\multicolumn{2}{c}{HD 171184} \\
Method                       &               
\multicolumn{2}{c}{\bz\ (G)} &
\multicolumn{2}{c}{\bz\ (G)} &
\multicolumn{2}{c}{\bz\ (G)} \\
\hline
(a) &$ -2506 $& 68 &$ -74 $& 39 &$ 125 $& 41 \\ 
(b) &$ -2622 $& 63 &$ -65 $& 39 &$ 139 $& 41 \\ 
(c) &$ -2549 $& 69 &$ -32 $& 47 &$ 150 $& 45 \\ 
(d) &$ -2512 $& 68 &$ -71 $& 39 &$ 166 $& 40 \\ 
(e) &$ -2452 $& 69 &$ -78 $& 39 &$ 127 $& 42 \\ 
(f) &$ -2496 $& 68 &$ -79 $& 39 &$ 148 $& 40 \\ 
(g) &$ -2508 $& 69 &$ -77 $& 39 &$ 127 $& 41 \\ 
(h) &$ -2453 $& 71 &$ -74 $& 40 &$ 110 $& 62 \\ 
(j) &$ -2499 $& 67 &$ -75 $& 39 &$ 117 $& 40 \\ 
(k) &$ -2811 $&101 &$-168 $& 47 &$ 171 $& 56 \\ 
(l) &$ -2406 $& 99 &$ 123 $& 77 &$  53 $& 62 \\ 
(m) &$ -3005 $&146 &$-185 $& 57 &$ 126 $& 71 \\ 
\hline
\end{tabular}
\end{center}
\noindent
Keys to the methods.
(a): as described at the end of Sect.~\ref{Sect_Details}; \ \
(b): same as (a), but with flat-fielding correction;\ \
(c): same as (a), but with optimal extraction;\ \
(d): same as (a), but using a 6 pixel aperture;\ \
(e): same as (a), but using a 18 pixel aperture;\ \
(f): same as (a), but without clipping;\ \
(g): same as (a), but without rectifying;\ \
(h): same as (a), with a 250 pixel wide filter;\ \
(j): same as (a), trimming only the first and the last 10 spectral points; \ \
(k): same as (a), but using Balmer lines only;\ \
(l): same as (a), but using metal lines.
(m): same as (k) (i.e., using Balmer lines only), but applying a 2\,pixel rebinning;\ \

\noindent
For all cases, we have computed error bars using Eq.~(\ref{Eq_External_Error}).
All observations were obtained with grism 600B and a 0.5\arcsec slit width.
Observing dates were 2003-02-08, 2004-04-17 and 2003-08-29 for HD~94660, HD~96441,
and HD~171184, respectively.

\end{table}
%%%%%%%%%%%%%%%%%%%%%%%%%%%%%%%%

The results discussed in the remainder of this paper are based on a
full reduction of the entire FORS1 ``fast mode'' data archive, making
the following choices: no flat-fielding is implemented, spectra are
extracted with average extraction using a 12 pixel radius aperture,
and rebinned at the dispersion value; the \pv\ profiles are rectified
using a 150 pixel wide Fourier filter, clipped with $k=3$ (including
also the two bins that are adjacent to the deviant one) as explained
in Sect.~\ref{Sect_Clipping}; the field is calculated from the full
spectrum, although the first 3\,\% and the last 3\,\% of spectral
points are ignored, and setting the Land\'e factor = 1 in proximity of
the H Balmer lines, and 1.25 everywhere else. This method is referred
to as method (a) in Table~\ref{Tab_Reductions}.  The choices made are
relatively conservative ones, tend to alter the actual data as little
as possible, and attempt to describe the uncertainties realistically.

We note that no rectification is implemented in the FORS pipeline,
since it would bring the aim of the software to a high-level of data
manipulation which is in fact outside of the scope of a common user
tool. Therefore, rather than using the final products of the pipeline,
we have always extracted the individual beams and treated them
individually according to the algorithm described above.

\subsection{Alternative techniques for field measurements}
%%%%%%%%%%%%%%%%%%%%%%%%%%%%%%%%%%%%%%%%%%%%%%%%%%%%%%%%%%%%%%%%%%%%%%%%%%%
\begin{figure}
\scalebox{0.45}{
\includegraphics*[0.8cm,6.5cm][22cm,25.5cm]{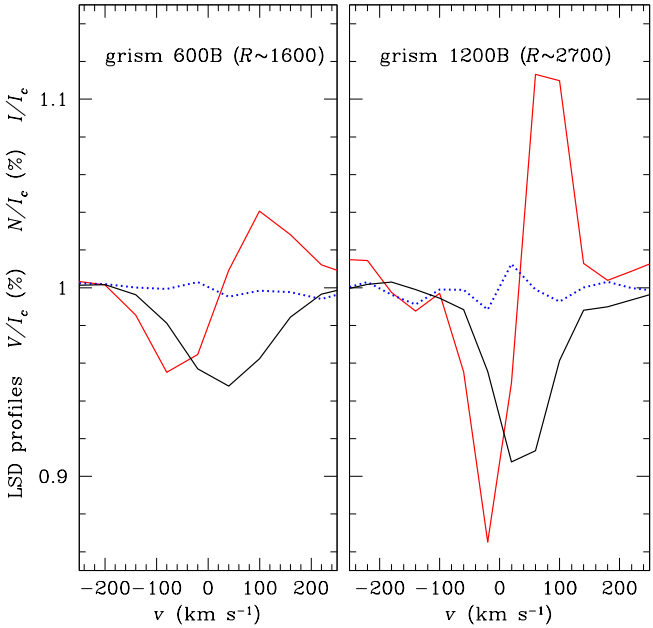}}
\caption{\label{Fig_LSD} LSD profiles for HD~94660 observed with
grism 600\,B (left panel) and 1200\,B (right panel). Red solid line:
Stokes $V/I_{\rm c}$; blue dotted line: $N/I_{\rm c}$; black solid line
Stokes $I/I_{\rm c}$. $V/I_{\rm c}$ and $N/I_{\rm c}$ are offset to +1\,\%
for display purpose.
}
\end{figure}
%%%%%%%%%%%%%%%%%%%%%%%%%%%%%%%%%%%%%%%%%%%%%%%%%%%%%%%%%%%%%%%%%%%%
Traditionally, two main techniques are used in the literature
to measure the mean longitudinal magnetic field. Measurements can be obtained from low
resolution spectropolarimetric data using the least-square technique
used in this paper, or (more commonly) from
high-resolution spectropolarimetric data treated with the Least-Square
Deconvolution (LSD) technique. In the latter case, the mean
longitudinal field is obtained from the measurement of the first-order
moment of LSD Stokes $V$ about the line centre, and of the equivalent
width of LSD Stokes $I$ -- both profiles normalised to the continuum
intensity \citep{Donetal97,Kocetal10}.

One may experiment with variations of these two techniques.
Equation~(\ref{Eq_Bz}) could be applied to high-resolution spectra, or
to their corresponding LSD profiles. LSD profiles themselves may be
obtained even from low-resolution data, allowing one to measure
\bz\ either with Eq.~(\ref{Eq_Bz}), or with the moment technique.  In
practice, none of these alternative methods has been adopted so far,
and it still has to be demonstrated that one can extract fully meaningful
LSD profiles from low-resolution spectra. While discussing alternative techniques
for \bz\ measurements is beyond the scope of this paper, we report on
our LSD experiments with three stars that were all observed with both
grisms 600\,B and 1200\,B: HD~358 (a non-magnetic star), HD~201601,
and HD~94660 (both magnetic). Using the moment technique, no field was
detected in the LSD profiles of HD~358, while the fields measured in HD~201601 and
HD~94660 were found to be $\sim 25$\,\% smaller (in absolute value) and more
dependent on the adopted grism, than those obtained with the technique
of Sect.~\ref{Sect_Field_Determinations}.  In all three cases, we
found no field detection in the null profiles, although for HD~94660
observed with grism 1200\,B, the null field value was = 2.4\,$\sigma$
(while with the least-square technique, the null field was consistent
with 0 within 1.6\,$\sigma$). We performed numerical simulations, which
confirmed that the \bz\ values obtained with the moment technique applied
to LSD profiles tend to underestimate
the actual field strength, at least for $\ga 1$\,kG field \citep[see
  also][]{Kocetal10}. Photon-noise error bars obtained with the two different
methods were found similar among themselves, but for the field obtained
from LSD profiles we found an additional large
uncertainty due to the continuum normalisation: the
\bz\ values measured with the moment technique applied to the LSD
profiles would change by more than 1\,$\sigma$ if spectra were
normalised according to different but still reasonable criteria. 
(We note that normalisation criteria of single order spectra are 
somewhat more arbitrary than in case of echelle spectra.)

In spite of these drawbacks, we note that in fast rotating stars, LSD
$V$ profiles may potentially reveal the signature of a magnetic field
even when its average longitudinal component is zero. Therefore,
applying LSD techniques to FORS data may be still of some interest.
Figure~\ref{Fig_LSD} shows the LSD profiles of HD~94660, observed with
grisms 600\,B and 1200\,B.

\section{Quality checks of the field measurements}\label{Sect_QC}
Inspection of the FORS1 archive data shows that between 2000 and 2009
about 1000 hours of telescope time were dedicated to acquiring stellar
circular spectropolarimetric data for magnetic field measurements.
Almost 600 hours of telescope time were granted in service mode, and 42
nights were granted in visitor mode.

\begin{figure}
\scalebox{0.45}{
\includegraphics*[0.5cm,5.8cm][22cm,25.5cm]{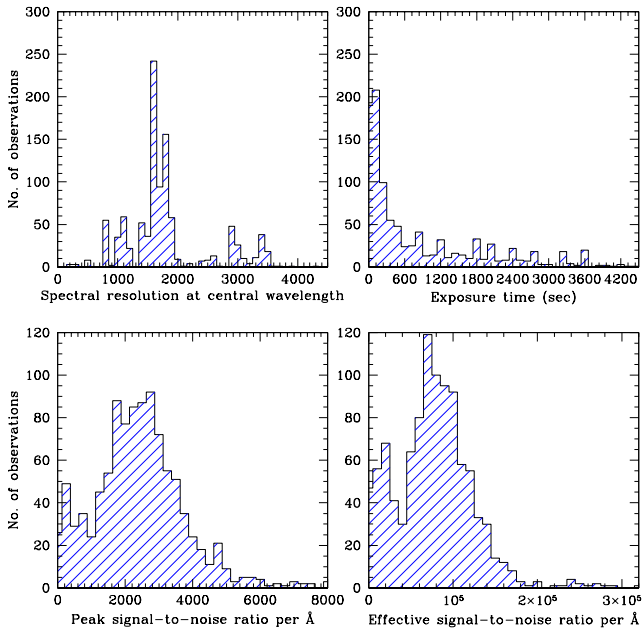}}
\caption{\label{Fig_Histograms} Some characteristics of the 
FORS spectropolarimetric observations retrieved in the archive.
}
\end{figure}

Figure~\ref{Fig_Histograms} gives a qualitative technical overview of
the nature of the observations, i.e., the distributions of the
spectral resolution, of the total exposure time per observing series,
of the peak signal-to-noise ratio per \AA, and the signal-to-noise
ratio integrated over all pixel bins used for the field determination
(when using the full spectrum).  The former two histograms demonstrate that
the FORS instrument was often used for very bright objects, pushing
the spectral resolution close to the limit fixed by sampling, and with
substantial slit losses. This suggests that some FORS1 users would
have rather benefitted from high-resolution spectropolarimetric facilities at smaller
telescopes. The latter two histograms show that most of the 
observations were obtained with a ultra-high signal-to-noise ratio, hence
FORS field measurements are generally close to the highest possible precision
obtainable with a reasonable use of the telescope time. The actual precision
of the FORS field measurement is the subject of this Section.

The discussion of Sect.~\ref{Sect_Details} shows that using different
reduction procedures to infer magnetic field strengths may lead to
significantly different results, in terms of both the mean longitudinal
field \bz\ and its associated uncertainty \sbz\ obtained from any
specific observation. In this Section we fix the data reduction
algorithm, and explore how the deduced fields and uncertainties still
lead to a number of inconsistencies. This discussion has two goals:
(1) to use what we know about some of the stars observed to try to
learn more about the characteristics of the FORS1 measurements, and
(2) to examine the reliability of various results, particularly
announced discoveries of new magnetic stars, based on FORS1
measurements.

\subsection{Internal tests}\label{Sect_Internal_Tests}
%%%%%%%%%%%%%%%%%%%%%%%%%%%%%%%%%%%%%%%%%%%%%%%%%%%%%%%%%%%%%%%%%%%%%%%%%%%
%%%%%%%%%%%%%%%%%%%%%%%%%%%%%%%%%%%%%%%%%%%%%%%%%%%%%%%%%%%%%%%%%%%%
In principle, calibration and testing of the measurements from any
spectropolarimeter should rely on calibration observations made on a
set of stars whose magnetic fields are well characterised by previous
work, particularly including some which are known to be non-magnetic
at the level of precision expected for the instrument.  Unfortunately,
such calibration measurements were not systematically obtained during
the working life of FORS1. Therefore we have searched the catalogue of
measurements for observations of stars with well known magnetic
characteristics to test the actual precision that has been achieved
with the FORS1 observations.

\subsubsection{Repeated observations of well-known magnetic stars}\label{Sect_Magnetic}
A dispersion test that has been very useful in determining how
realistic the measurement uncertainties are of other instruments such
as the MuSiCoS spectropolarimeter \citep[e.g.,][]{Wadetal00} is
to examine the measurements of \bz\ as a function of the star's
rotation phase around the mean variation.  Thus, for example, 17
measures of \bz\ of $\beta$~CrB with MuSiCoS with uncertainties of 20
to 30~G were fully consistent with the mean curve at signal to noise
ratios of up to 30:1. This kind of instrument monitoring was not
performed with the FORS1 instrument, and we suggest it should be
implemented in the FORS2 calibration plan. A somewhat useful test was
performed by some users who repeatedly observed the magnetic star
HD\,94660. The star's rotation period is not known, but is probably
longer than two years. The star's longitudinal field appears nearly (but
not perfectly) constant and about $-2$\,kG, thus spectropolarimetric
observations of HD\,94660 allow one to assess if the
polarimetric optics are at least approximately aligned.

\subsubsection{Repeated observations of non-magnetic stars}\label{Sect_Non_Magnetic}
One particular test that should be carried out frequently as part of
routine calibration of any spectropolarimeter used for magnetic
measurements is the observation of stars known to be non-magnetic, or
at least to be non-magnetic to within rather small uncertainties (a
few G or tens of G). Such measurements are a particularly powerful
test of the uncertainties computed for individual observations, since
the actual field value expected (zero) is known with high precision.
An example of such a test was the monitoring of $\delta$~Eri by
\citet{Donetal97} with the SEMPOL instrument at the Anglo-Australian
Telescope (AAT). No main sequence stars known with high precision to be non-magnetic
have been \textit{deliberately} observed with FORS1. However, some of
the peculiar A stars of the HgMn type, and in particular $\alpha$~And
= HD~358, have been repeatedly observed, probably because they were suspected to
be magnetic. Numerous investigations of magnetism in such stars \citep[e.g.][]
{BorLan80,Glagetal85,Shoretal02,Aurietal10,Makagetal11} have found no convincing evidence for fields in
this class of stars with uncertainties often in the range of a few G
up to a few tens of G.  Furthermore, $\alpha$~And has been
specifically investigated with the MuSiCoS (13 observations, typical
$\sigma \sim 30 - 60$~G) and ESPaDOnS (5 observations, typical $\sigma
\sim 6 - 19$~G) spectropolarimeters without any significant detection
of a non-zero longitudinal field or of any Stokes $V$ signature
indicative of a non-zero field \citep{Wadetal06}.

%%%%%%%%%%%%%%%%%%%%%%%%%%%%%%%%%%%%%%%%%%%%%%%%%%%%%%%%%%%%%%%%%%%%%%%%%%%%%%%%%%%%%%%
\begin{table}
\caption{\bz\ measurements of $\alpha$~And = HD~358}
\centering
\begin{tabular}{cr@{$\pm$\,}lr@{$\pm$\,}lr@{$\pm$\,}l}
\hline\hline  
  Date      & \multicolumn{2}{c}{\citet{Hubetal06b}}& \multicolumn{4}{c}{This work}\\
yyyy-mm-dd  & \multicolumn{2}{c}{\bz\ (G)}          & \multicolumn{2}{c}{\bz\ (G)}&\multicolumn{2}{c}{\nz\ (G)}\\
\hline
2003-09-28  &\ \ \ \ \ \ $-261$ &  73      &$ -296$&  121 &$ -51$ & 114 \\
2003-11-20  &\ \ \ \ \ \     12 &  82      &$  160$&  104 &$-143$ & 121 \\
2005-05-29  &\ \ \ \ \ \ $-109$ &  49      &$  -40$&  113 &$ 216$ & 103 \\
2005-09-16  &\ \ \ \ \ \  $-73$ &  20      &$   23$&  33  &$  12$ &  31 \\
2005-09-17  &\ \ \ \ \ \  $-30$ &  30      &$  -12$&  39  &$ -30$ &  34 \\
2005-09-25  &\ \ \ \ \ \ $-108$ &  23      &$  -36$&  37  &$ -26$ &  36 \\
2007-11-28  &\multicolumn{2}{c}{}&$  107$&  34  &$ 111$ &  31 \\
2007-11-29  &\multicolumn{2}{c}{}&$  -20$&  24  &$ -29$ &  20 \\
\hline
\label{Tab_Alpha_And}
\end{tabular}
\end{table}
%%%%%%%%%%%%%%%%%%%%%%%%%%%%%%%%%%%%%%%%%%%%%%%%%%%%%%%%%%%%%%%%%%%%%%%%%%%%%%%%%%%%%%%

The field of $\alpha$ And has been measured eight times with FORS1.  The
first six observations have been published \citep{Hubetal06b}.
Table~\ref{Tab_Alpha_And} lists the published \bz\ measurements and the
\bz\ values of the same observations resulting from our reduction.

We note several interesting features of the data in this table. First,
our new reduction invariably gives somewhat larger measurement
uncertainties than the published data. Secondly, the published data
report three field detections at the level of $3\,\sigma$ or higher; our
new reduction finds only a single detection at the $3\sigma$ level,
and it is {\it not} one of the three initially reported $3\,\sigma$
measurements. Furthermore, the null field \nz\ (i.e., the magnetic
field calculated using the null profiles instead of the \pv\ profiles)
is also detected once at about the $3\,\sigma$ level. Thus, our new results,
independently of any external evidence, make the detection of a
longitudinal field in this star from FORS1 data very uncertain.  The
quite substantial differences between the field values published
previously and those we have measured are probably not due to errors
of reduction, but to the sensitivity of the results to the specific
reduction method used to obtain the \bz\ value.

If we now assume that the more precise measurements reported by
\citet{Wadetal06} establish that no longitudinal field is
present in $\alpha$~And at the level of 20 or 30\,G, so that
effectively we assume that $\alpha$ And is a null standard, we may
regard the results of Table~\ref{Tab_Alpha_And} as providing useful
information about the behaviour of the \bz\ values and associated
uncertainties obtained from FORS1 measurements. In particular, both
\bz\ and \nz\ values scatter from zero by more than is expected from
the standard deviation of their individual measurements.  Our results
suggests that even the ``external'' error of
Eq.~(\ref{Eq_External_Error}), which is still ``internal'' to a single
measurement, leads to occasional underestimates of the actual
\bz\ error bars.

\subsubsection{Time series}\label{Sect_Time_Series}
A further test of the stability of measured field values, and of the
associated uncertainties, can be obtained by studying the statistical
properties of a long series of magnetic observations of the same star
taken during a single night. One example is the study by
\citet{Hubetal04b}, who observed six known cool magnetic rapidly
oscillating Ap stars (roAp stars). These observations had a cadence of
the order of one frame per minute, and were typically continued for
one to a few hours over ten or twenty pulsation cycles. The
observations were intended to search for possible periodic variations
in the magnetic field \bz\ at known pulsation periods.  According to
the published result of this study, no short-period field variations
were reliably detected. If we make the assumption that during the
short time interval in which each star was observed, the field was in
fact constant, we can compare the standard deviation of the values of
$\bz_i$\ about their mean value $\widehat{\bz}$ with the computed
uncertainties $\sbz^{(i)}$. If this standard deviation (the ``external
error'') is similar to the published values of $\sbz^{(i)}$ (``internal
errors''), then the comparision supports the correctness of the
internal errors.  Excess standard deviation indicates problems with
deduced values of \sbz. 

This comparison was carried out by the team who did the observations
\citep[see Table~9 in][]{Hubetal04b}. Of the six stars reported,
\textit{only two had external errors close to the internal ones}. The
other four stars all showed excess scatter, up to an extreme
external error of more than three times the internal errors for
HD\,201601. One interpretation of this excess, of course, would be
that these roAp stars were actually displaying pulsational field
variations, but in only one case could a frequency be identified that
is similar to one of the known pulsation frequencies of the roAp star,
and a second observing run on this star (HD\,101065) revealed no
signal at that frequency \citep{Hubetal04b}. Furthermore, HD\,101065
is one of the two stars observed for which external and internal
errors are very similar in magnitude; none of the stars with a
significant discrepancy between internal and external errors showed
significant variations at known pulsation periods. Thus it seems far
more likely that the difference between internal and external errors
reveals something about FORS1 and/or the reduction methods used for
spectropolarimetric observations.

We have repeated the dispersion analysis for these stars, using our
new reductions.\footnote{One particular feature of the newly reduced
  data is the change in sign of the data for HD\,83368 compared to the
  earlier publication by \citet{Hubetal04b}.  This point has been
  carefully investigated, and we believe that the signs in
  \cite{Hubetal04b} are incorrect for all the data for this star, and
  should be changed to positive signs.}  However, rather than using
\bz, for this kind of analysis we have preferred to use the mean
longitudinal field \nz\ from the null profiles. The reason is that
Eq.~(\ref{Eq_Bz}) is strictly valid only in the weak-field regime, and
for non blended spectral lines. These conditions are not met in the
observed roAp stars, hence it is more likely that the null field \nz\ has a more
linear response to the photon noise than the magnetic field \bz.
We have proceeded in the following way. 
From all $M$ frames of a given observing series
we have selected $4 \times n$ frames, where $n$ is the integer part of $M/4$,
and calculated $n$ $\nz^{(i)} \pm \snz^{(i)}$\ values.
Then we have calculated:\\
the weighted mean of the $\nz^{(i)}$ value
%%%%%%%%%%%%%%%%%%%%%%%%%%%%%%%%%%%%%%%%%%%%%%%%%%%%%%%%%
\begin{equation}
\widehat{\nz} = \frac{\sum_i^n \nz^{(i)}/\left(\snz^{(i)}\right)^2}{\sum_i^n 1/\left(\snz^{(i)}\right)^2} \; ,
\end{equation}
%%%%%%%%%%%%%%%%%%%%%%%%%%%%%%%%%%%%%%%%%%%%%%%%%%%%%%%%%
the average ``internal'' standard deviation
%%%%%%%%%%%%%%%%%%%%%%%%%%%%%%%%%%%%%%%%%%%%%%%%%%%%%%%%%
\begin{equation}
\widehat{\snz} = \frac{1}{n}\ \sum_i^n \snz^{(i)} \; ,
\end{equation}
%%%%%%%%%%%%%%%%%%%%%%%%%%%%%%%%%%%%%%%%%%%%%%%%%%%%%%%%%
the ``external'' standard deviation
%%%%%%%%%%%%%%%%%%%%%%%%%%%%%%%%%%%%%%%%%%%%%%%%%%%%%%%%%
\begin{equation}
\snz' = \left(\frac{\sum_i^n (\nz^{(i)} - \widehat{\nz})^2}{n-1}\right)^{1/2} \; ,
\end{equation}
%%%%%%%%%%%%%%%%%%%%%%%%%%%%%%%%%%%%%%%%%%%%%%%%%%%%%%%%%
and error of the mean
%%%%%%%%%%%%%%%%%%%%%%%%%%%%%%%%%%%%%%%%%%%%%%%%%%%%%%%%%
\begin{equation}
\epsilon = \left(\frac{\sum_i^n (\nz^{(i)} - \widehat{\nz})^2}{n(n-1)}\right)^{1/2}  = \frac{\snz'}{n^{1/2}} \; .
\end{equation}
%%%%%%%%%%%%%%%%%%%%%%%%%%%%%%%%%%%%%%%%%%%%%%%%%%%%%%%%%
We have also combined all $4n$ frames and obtained a $\bz^{[n]} \pm \sbz^{[n]}$ value
for each star.

%%%%%%%%%%%%%%%%%%%%%%%%%%%%%%%%%%%%%%%%%%%%%%%%%%%%%%%%%%%%%%%%%%%%%%%%%%%%%%%
\begin{table}
\begin{center}
\caption{\label{Table_roAp} Null field values and their error bars for six roAp stars}
\centering
\begin{tabular}{lrrrrrrr}
\hline\hline
Star      & $n$ & $\widehat{\nz}$ & $\widehat{\snz}$ & $\snz'$& $\nz^{[n]}$ & $\snz^{[n]}$  &    $\epsilon$\\
\hline                                                                                                          
HD 83368  & 31  & $    -5       $ & $     44       $ & $     67   $ &  $ -1    $ & $  8   $  &   $    12   $ \\
HD 101065 & 16  & $    -9       $ & $     40       $ & $     37   $ &  $ -6    $ & $ 11   $  &   $     9   $ \\
HD 128898 & 14  & $    11       $ & $     53       $ & $    101   $ &  $ -2    $ & $ 15   $  &   $    27   $ \\
HD 137949 & 16  & $     8       $ & $     41       $ & $     70   $ &  $ 17    $ & $ 11   $  &   $    17   $ \\
HD 201601 & 17  & $    86       $ & $     80       $ & $    130   $ &  $ 72    $ & $ 19   $  &   $    32   $ \\
HD 217522 & 50  & $    39       $ & $    110       $ & $    190   $ &  $ 23    $ & $ 16   $  &   $    27   $ \\
HD 101065 & 36  & $     4       $ & $     45       $ & $     58   $ &  $ -6    $ & $  8   $  &   $    10   $ \\
\hline
\end{tabular}
\end{center}
\end{table}
%%%%%%%%%%%%%%%%%%%%%%%%%%%%%%%%%%%%%%%%%%%%%%%%%%%%%%%%%%%%%%%%%%%%%%%%%%%%%%%
Table~\ref{Table_roAp} reports these quantities for each observed
star. While there is good agreement between the $\widehat{\nz}$
and $\nz^{[n]}$ values, in all cases but one we found $\snz' >
\widehat{\snz}$ and, consistently, $ \epsilon > \snz^{[n]}$. Our
data sets appear to reveal excess noise over what would be expected
from the internal uncertainties. Excess noise arises irregularly: it
increases the standard deviation by a factor varying from 1 to about
2, qualitatively suggesting that the standard errors of measurement of
\bz\ from internal dispersions within each measurement are
underestimates of the true uncertainties.

\subsubsection{The distribution of the magnetic field from the null profiles}\label{Sect_Null_Fields}
%%%%%%%%%%%%%%%%%%%%%%%%%%%%%%%%%%%%%%%%%%%%%%%%%%%%%%%%%%%%%%%%%%%%%%%%%%%
\begin{figure}
\scalebox{0.45}{
\includegraphics*[0.5cm,4.8cm][22cm,25.5cm]{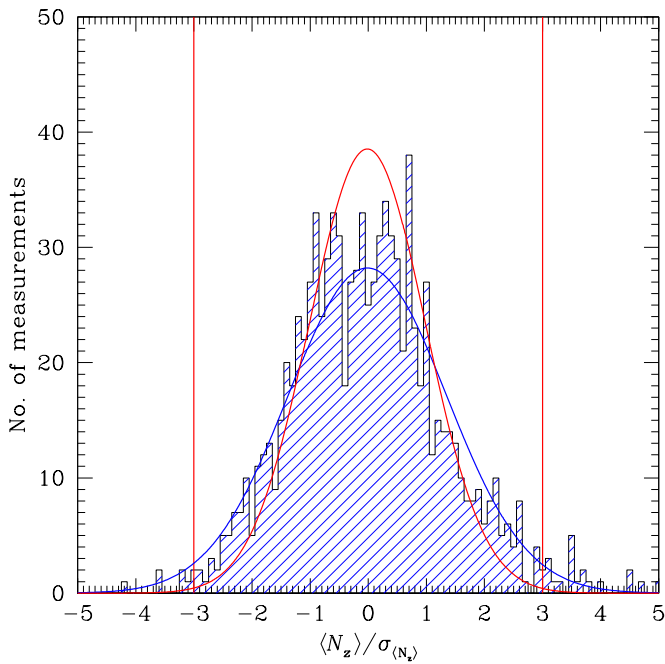}}
\caption{\label{Fig_Histonull} 
Histogram of the \protect\nz\ values calculated from Balmer lines, and
normalised to the error bars \protect\snz. The red solid lines show the
gaussian with $\sigma=1$, and the blue solid line a gaussian with
$\sigma=1.365$, both with the same areas as the observed
distribution. The vertical lines mark the limits at -3 and 3.
}
\end{figure}
%%%%%%%%%%%%%%%%%%%%%%%%%%%%%%%%%%%%%%%%%%%%%%%%%%%%%%%%%%%%%%%%%%%%
A final internal test is to compute the distribution over the entire
FORS1 sample of field measurements of the {\it null field values}
normalized by their error bars. 

The null profiles should represent measurements of zero fields, as the
combinations of different waveplate settings used are chosen so that
the real polarisation signal cancels out, to the precision of the
measurements. If the field values measured from the null profiles have
a distribution determined by photon noise only, this distribution
should closely resemble to a gaussian distribution with $\sigma=1$.
Figure~\ref{Fig_Histonull} shows that there is a significant excess of
$\nz/\snz$ points outside of the interval $[-3,3]$. The standard
deviation of the $\nz/\snz$ is 1.37, 1.20, 1.36 for the Balmer lines,
metal lines, and the full spectrum, respectively. 

\subsection{Possible explanations for the internal inconsistencies}\label{Sect_Discussion}
The results obtained for non magnetic stars, and the analysis of the time
series and of the distribution of the null
profiles, suggest that error bars, even when calculated with
Eq.~(\ref{Eq_External_Error}) -- which are generally higher than those
obtained Eq.~(\ref{Eq_Internal_Error}), are underestimated.  Our
conclusion is that Poisson noise is not the only source of
uncertainty of the field measurements. We have already seen that an
additional source of uncertainty is introduced by the choice of
the algorithm for data reduction, but the results of
Sect.~\ref{Sect_Time_Series} and \ref{Sect_Null_Fields} were obtained
consistently using the same algorithm. Therefore we suggest that some
physical, probably instrumental, limit is hit, especially when
measurements are obtained with ultra-high signal-to-noise ratio.

Possible reasons that could contribute to explaining this phenomenon are:

\noindent
\textit{i)} Undetected problems with the raw data, such as saturated
frames, or various kinds of errors in the organisation of the
scientific frames and their calibrations, which range from the awkward
mixing up of frames of different stars, to more difficult to detect
hardware failures which could be responsible for an erroneous
recording of the position angle of the retarder waveplate in the
fits-headers (these problems are known to have occurred at some time
in the FORS1 data).

\noindent
\textit{ii)} Change of the line profiles during the observations for
reasons intrinsically due to the source (e.g., stellar pulsations).

\noindent
\textit{iii)} Changes of the radial velocity of the target occurring
during the observing series.

\noindent
\textit{iv)} Small instrument flexures occurring due to instrument
motion during the observing series.

\noindent
\textit{v)} For series of very short exposures in good seeing,
changes in the mean position of the star in the slit from one exposure
to the next, with corresponding changes in the mean wavelengths
associated with each pixel. We note that many observations were obtained 
with exposure time of 0.25\,sec per frame.
%%%%%%%%%%%%%%%%%%%%%%%%%%%%%%%%%%%%%%%%%%%%%%%%%%%%%%%%%%%%%%%%%%%%%%%%%%%
\begin{figure}
\scalebox{0.45}{
\includegraphics*[0.5cm,4.8cm][22cm,25.5cm]{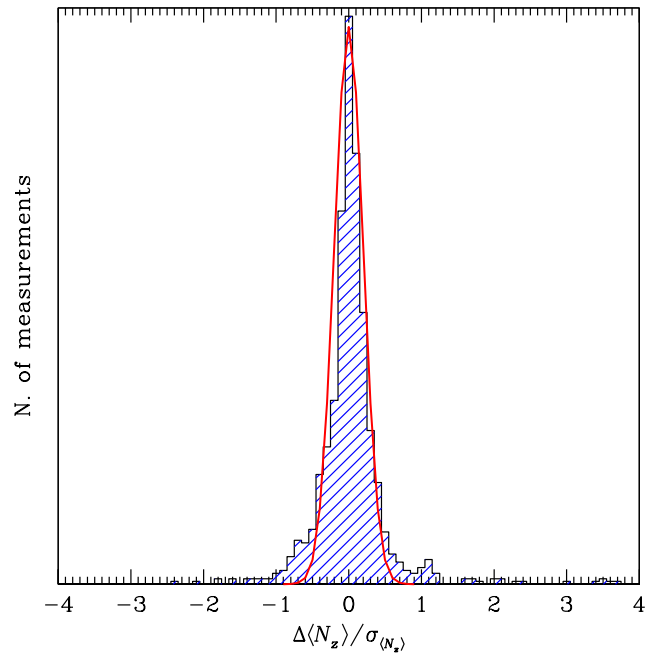}}
\caption{\label{Fig_Flexures} The distribution of the differences between 
the null field values measured after flexure simulation
and their original values, normalised to the error bars, compared to a Gaussian
curve centred about zero and with $\sigma=0.2$. Null field values were
calculated from the H Balmer lines.
}
\end{figure}
%%%%%%%%%%%%%%%%%%%%%%%%%%%%%%%%%%%%%%%%%%%%%%%%%%%%%%%%%%%%%%%%%%%%

We have clearly tried to minimise problems similar to those described in
\textit{i)}, and issues \textit{ii)} and \textit{iii)} can be
investigated case by case. In this regard, it is interesting to note
that many dubious field detections appear in slowly pulsating stars
(see Sect.~\ref{Sect_Beta_SPB}).

In the case of FORS1 and FORS2, instrument flexures are kept at a
minimum.  However, we must note that ultra-high precision polarimetric
measurements can be performed only with an extremely stable
instrument, and it is for that reason that the ESPaDOnS and NARVAL
instruments are sitting in thermally and mechanically isolated
benches. The question is what level of stability the Cassegrain
mounted FORS instruments can guarantee, and how possible instrument
flexures would impact the \bz\ and \nz\ measurements. This does not
have a simple answer. Numerical simulations on a single spectral line
show that a drift in wavelength equal in both beams (e.g., due to a
star's radial velocity change) has a much stronger impact on the null
profiles than on the Stokes profiles. A wavelength offset
that acts differentially on the two beams, and that depends on the
position of the retarder waveplate, has a larger impact on
the Stokes profiles than on the null profiles\footnote{This case is
  illustrated in \citet{Bagetal09}, who also
  claimed that the null profiles obtained with the difference method
  are more sensitive to differential wavelength offsets than the null
  profiles obtained with the ratio method. In fact, we found that this
  claim is wrong. Both ways to calculate the null profiles are
  equivalent, and in the cases (c) and (d) presented in their Fig.~6,
  both null profiles obtained with the ratio and the different methods
  are zero.}.
We note also that instrument flexures affect not only the Stokes and
the null profiles, but also the normalised derivative of the flux
\citep[which is calculated as explained in Sect.~4.2 of][]{Bagetal02},
and, finally, that the impact of all these effects on \bz\ and
\nz\ values depends also on line profile and blending.

In an attempt to roughly evaluate the potential consequences of
instrument flexures, we have performed the following test. In the
data reduction process, we have numerically simulated a drift
in wavelength continuously and homogeneously acting between the first
and the last exposure. The total amplitude of the drift (i.e.,
calculated between the last exposure and the first exposure), was
equal to a quarter of the spectral bin. Then we have computed the
difference $\Delta \nz$ between the \nz\ value so obtained, and the
value obtained without introducing any offset. We have repeated this
simulation for all FORS archive datasets, and calculated the
distribution of the ratio $\Delta \nz / \snz$. Figure~\ref{Fig_Flexures}
displays the results obtained from H Balmer lines. This histogram
shows that even a nearly negligible offset during the observations may
produce changes in the \nz\ determinations comparable to the error
bars. For the narrower metal lines, the consequences of flexures are
more pronounced, and the \nz\ offsets being twice larger than when
calculated from H Balmer lines. Qualitatively similar results was
retrieved for the \bz\ values. However, we should note that for a
certain dataset, a given simulated drift does not necessarily produce
an offset with identical magnitude for both the \bz\ and
\nz\ values. This suggests that, at least for small flexures, the
warning coming from the \nz\ detections has a \textit{statistical}
relevance. For individual cases, a detection in \nz\ does not
necessarily invalidate a \bz\ detection in the same
dataset. Viceversa, a \bz\ detection may be spurious even if there is
no detection in the null profile.

According to the FORS instrument manual, in imaging mode the
instrument is stable within 0.2 pixel during 1\,hour
exposure. However, Fig.~2.4 of that manual shows that flexures as
large as a quarter of a pixel are not uncommon if the instrument
rotator rotates by $10-20^\circ$.  Furthermore, the instrument manual
refers to a situation in imaging mode; the additional optical elements
introduced in spectro-polarimetric mode (PMOS mask, slit mask,
Wollaston prism, retarder waveplate, grism, and possibly a filter)
certainly cannot improve the instrument stability. While we conclude
that instrument flexures may explain at least in part why FORS field
measurement uncertainties seem to be underestimated, we must also
point out that we were unable to find a correlation between
\nz/\snz\ values and the airmass.

\section{The incidence of magnetic fields in various classes of stars}\label{Sect_Incidence}
Our conclusion that the error bars computed from photon statistics
underestimate by at least 30\,\%--40\,\% the real uncertainties of
field measurements has significant implications for the body of data
obtained from FORS1, in particular for those measurements which have
been used to announce new discoveries of a magnetic field at a
relatively low significance level. Because of problems that spurious
discovery announcements can cause, in the rest of this paper we
concentrate on those stars in which a magnetic field was detected with
FORS1 at $\la 6\,\sigma$ level, and present a detailed comparison of
our new results with previously announced discoveries of magnetic
fields.  A homogeneous and comprehensive catalogue of all FORS1
measurements will be published in a forthcoming paper.

The results of our comparison are summarised in
Table~\ref{Tab_Discoveries}.  The stars are grouped into families of
various kinds, identified by bold-face titles such as ``HgMn stars'',
``Be stars'', etc.  The first two columns in this table list the star
name, and the mean MJD of the observation as computed for the frames
used in our data reductions (a blank in one of these two columns means
that the line refers to a second published value of the same data as
appears immediately above). The third and fourth columns give the
value of \bz\ and its uncertainty as published, and a code denoting
the reference where the field measurement was published. The fifth
column reports the \bz\ value and its uncertainty obtained from our
adopted reduction process. The sixth column includes a code character
describing our conclusion on whether a field is detected in the star:
D means that the star is definitly magnetic; P means that a field is
possibly present, but that the currently available measurements do not
provide firm evidence for that; N means that none of the measurements
shows evidence of the presence of a magnetic field.  Note that a few
of the entries in Table~\ref{Tab_Discoveries} have MJD values that
differ from those published, mostly by 0.5 day. Our values are
computed directly and uniformly from the data in the fits headers of
the observations, and should replace the published values where these
differ.

Note that Table~\ref{Tab_Discoveries} does not include the
``marginal'' fields detections in Ap and Bp stars -- most of which are
reported by \citet{Hubetal04b}, \citet{Bagetal06}, \citet{Hubetal06b}
and \citet{KocBag06}. It is already clearly demonstrated that magnetic
fields are detected in essentially every Ap star if high enough
precision is reached \citep{Auretal07}. Therefore, the reliability of
the detection of weak fields reported in these papers will not greatly
change what we know about the occurrence of magnetic Ap and Bp stars.
Accordingly, we postpone a discussion of these data to our forthcoming
paper.

The uncertainty of our field measurements of
Table~\ref{Tab_Discoveries} is calculated with
Eq.~(\ref{Eq_External_Error}), which is still an underestimate of the
real error bar. Unfortunately, the source of errors that are not due
to photon noise cannot be estimated for the individual cases. Although
one may be strongly tempted to multiply all error estimates by a
factor $\sim 1.5$, the correct way to proceed is to remember that
3\,$\sigma$ level is not a safe detection threshold.  For that reason,
the discussion presented in this Section will also rely on checking
whether a field detection has been repeatedly confirmed with numerous
FORS1 measurements, and on corroborating or conflicting field
measurements made with other spectropolarimeters.

We found that several important FORS1 discoveries announced in the
past are simply not confirmed by our new data reduction, \textit{even without
taking into account a possible underestimate of our error bar}. In many
cases, in contrast to the originally published result, the absolute
value of our new field measurement does not exceed three times its
associated $\sigma_{\bz}$ value estimated with
Eq.~(\ref{Eq_External_Error}). We have two possible general
explanations for this finding. The first one is based on the general
considerations of Sect.~\ref{Sect_Details} that different data
reduction methods lead to slightly different results. Also, in some
cases we have removed some frames (e.g., because of saturation), or
used a slightly different frame selection than used for the original
publication. In an attempt to keep this aspect under control, we have
compiled a very large catalogue that includes, for each observation,
all \bz\ values obtained with several different data reduction methods
(e.g, without rectifying Stokes profiles, or using optimal extraction
instead of average extraction). In the following discussion we will
explicitly comment on those cases in which one or more methods for
data reduction alternative to the one used to compile
Table~\ref{Tab_Discoveries} would lead to a qualitatively different
result. These cases are comparatively rare. A second explanation of
the discrepancies between our results and those published in the
literature is that our error bars are generally larger than those
obtained in previous works. In the following we comment in detail on
our new results.

\subsection{Am, HgMn and PGa stars}
The HgMn peculiar late B stars have already been mentioned in
Sect.~\ref{Sect_Non_Magnetic}, where we concluded that the
overwhelming bulk of evidence is against the presence of fields having
\bz\ values of more than a couple of tens of
G. Table~\ref{Tab_Discoveries} contains six measures claimed as field
detections of four such stars by \citet{Hubetal06b}. The new
reductions have increased the standard errors of all the measurements
by factors of 1.3 or more, in two cases by much more. None of the
measurements now shows a field at the $3\,\sigma$ level, and three of
the five measures that were reported to be over 100\,G and detected at
even $4\,\sigma$ or more are now only some tens of G, and at roughly
the $1\,\sigma$ level. In contrast, two of the large values are still
relatively large, but are no longer significant at the $3\,\sigma$
level, and may represent the occasional spurious large field values
that appear in a few percent of FORS1 field measurements.

We conclude that the FORS1 data do not alter the general non-detection
of magnetic fields in HgMn stars. There is still no significant
evidence of fields in such stars. 

The PGa stars are considered to be the extension of the HgMn
phenomenon to higher effective temperatures. A survey of several such
stars some years ago \citep{Boretal83} revealed no fields, with
uncertainties of hundreds of G. The large FORS1 survey reported by
\citet{Hubetal06b} included two PGa stars observed with much
smaller uncertainties. One of these stars, HD~19400, was reported
to have a field. Our re-reduction of these data show no significant
field, and we consider this detection to be spurious. 

The Am (metallic-line) stars are considered to be analogues of
  the HgMn stars at lower $T_{\rm eff}$. There have been only a few
  observations of such stars with FORS1, and no claims of field
  detections based on these observations have been made.

\subsection{Classical Be stars}\label{Sect_Be_Stars}
By classical Be stars we mean those B stars which show evidence (at
least some of the time) of centrifugally supported, slowly evolving,
fairly massive disks. For many years some theoreticians have argued
that magnetic fields may play an important role in the creation and/or
maintenance of the circumstellar disks. However, there have not been
very many serious searches for magnetic fields in such stars. A search
by \citet{Baretal85} with uncertainties in the range
of 50 -- 250~G did not reveal any fields. 

On the basis of data from the MuSiCoS spectropolarimeter,
\citet{Henetal00}, \citet{Neinetal01} and \citet{Donetal01} have
reported the detection of a weak field in HD\,205021 = $\beta$\,Cep, a
double system composed of a Be star and a pulsating star  -- the latter
being the proto-type of $\beta$\,Cep pulsating
stars. \citet{Schnetal06} have shown that the primary star of the
$\beta$~Cep system is the pulsating and magnetic star, while the Be
star is a non magnetic companion that can be separated with speckle
interferometry.  Thus the detected field is not connected with the Be
phenomenon.  \citet{Neinetal03a} reported a magnetic field in the
classical Be star $\omega$~Ori, but it is not clear that the data
provide a convincing case for the presence of a field.
\citet{Siletal09} have observed the two classical Be stars HD~148184
and HD~181615 several times each with ESPaDOnS, with uncertainties of
order 100~G and 10~G respectively, but found neither significantly
non-zero values of \bz, nor signficant polarisation signatures. Thus
from the limited data available from instruments other than FORS1,
there is no strong evidence for magnetic fields in any classical Be
star.

The largest recent surveys for fields in classical Be stars have been
carried out using FORS1 by \citet{Hubetal07b} and \citet{Hubetal09b}
\citep[summarised by][]{Yudietal11}, both searching for magnetism in
field Be stars, and by \citet{McSwain08}, who studied normal, chemically
peculiar, and emission-line B stars in the young open cluster
NGC~3766. Between them, these surveys have included more than 40 Be
and possible Be stars, of which 10 stars were reported to have weak
but significant magnetic fields. The reports by \citet{Hubetal07b} and
\citet{Hubetal09b} suggest that weak ($\sim100$\,G) fields may be
ubiquitous in Be stars, a result that would have important
consequences for modelling the Be phenomenon.

Note that there is some doubt as whether all of the stars in these
surveys are Be stars. For example, NGC~3766 MG 94, 111, and 176 are
listed as a He strong star, normal star, and normal star by
\citet{McSwain08}, but as B stars with emission lines in Table~2 of
\citet{Hubetal09b}. In Table~\ref{Tab_Discoveries} we followed the
classication by \citet{McSwain08}. Furthermore, there are differences
among these publications as to which stars show field detections;
\citet{McSwain08} did not find a significant field in the normal B
stars NGC~3766 MG 111 or NGC~3766 MG 176, nor in the Be star NGC~3766 MG 200,
all of which are reported, using the same data, as new detections by
\citet{Hubetal09b}.

When we compare the new reductions of these observations with the
published ones, we see that only one of the reported \bz\ values that
differ from zero by more than $3\,\sigma$ are still significantly
non-zero. In part this is because most (but not all) of the
measurement uncertainties resulting from our re-reductions are larger
than those originally reported, and in part because most of the new
\bz\ values are also smaller (in absolute value) than originally
reported. The only remaining apparently significant detection is in
HD~181615; however, the photon-noise error of the measurement that
still shows field detection in our reduction is certainly very small,
and the presence of a field of the claimed size is clearly
contradicted by the three negative observations of \citet{Siletal09}.

There is a huge discrepancy between the uncertainty of the
published field detections for HD~148184 and our uncertainties in
Table~\ref{Tab_Discoveries}. These spectra are heavily polluted
with emission lines, and we were unable to reproduce the very tiny
field uncertainties published by \citet{Hubetal07b}. However, we must
note that the quality of our reduced data corresponding to the
second detection is particularly poor, probably due to image quality. 
We note that the published detections are contradicted by four negative
observations of \citet{Siletal09} with uncertainties mostly of about
70~G.

Our reductions using the ESO pipeline give apparently significant
field detections for an observation of HD~56014 and two observations
of HD~209409, in which the original observers did not find report any
significant fields (for that reason, the corresponding entries are not
included in Table~\ref{Tab_Discoveries}). We also obtained a
$15\,\sigma$ detection (!) in HD~224686, in which the original
observers measured $\bz = 74 \pm 24$\,G.  All these data were obtained
in the same two nights, MJD=54432 and 54433.  We are very sceptical of
our reductions, since they all show a significant signal in the bluest
Balmer lines both in the \pv\ and in the \nv\ profiles, which we
believe the symptom of a not yet understood problem with calibrations
or with our data reduction. In Table~\ref{Tab_Discoveries}, for
HD~224686, instead of our field estimate, we left an empty space.

Our global conclusion is that most if not all of the detections
reported in Be stars by \citet{Hubetal07b} and \citet{Hubetal09b}, as
well as the marginal detection reported by \citet{McSwain08}, are
probably spurious, and that magnetic fields much above 100~G rarely if
ever occur in classical Be stars.

\subsection{Herbig AeBe stars}
The Herbig AeBe stars are pre-main sequence stars which are in the
final stages of contraction, and which are destined to become main
sequence A- or B-type stars. These stars are identified by their locations
in regions of current or very recent star formation, by the presence
of optical emission lines (especially H$\alpha$) produced by
circumstellar or accreting material, and by the presence of an
infrared excess due to cool circumstellar material not yet accreted
onto the star. The stars themselves have spectral types A or B.

The first discovery of a field in a Herbig Ae star, HD~104237, was
reported by \citet{Donetal97} from observations at the
AAT with an experimental spectropolarimeter.
The star was observed again with SEMPOL at the AAT by E.\ Alecian
(private communication) who confirms the magnetic detection previously
reported by \citet{Donetal97}.

Based on FORS1 observations, the discovery of a magnetic field
in the Herbig Ae star HD~139614 was reported by \citet{Hubetal04a},
who measured $\bz = -450 \pm 93$\,G. \citet{Wadetal05} re-analysed
the FORS1 dataset acquired by \citet{Hubetal04a}, and obtained the
more marginal result
$\bz = -150\pm50$\,G. \citet{Wadetal05} reported also that ESPaDOnS
data did not show any presence of a magnetic field. A second detection
in HD~139614 ($\bz=-116\pm34$\,G) was claimed by
\citet{Hubetal06c}, who also confirmed their previous measurement of -450\,G.
Later, \citet{Hubetal07a} revised both FORS1 observations, and reported
$\bz -112 \pm 36$\,G and $\bz=-93\pm14$\,G, for the first and the
second measurement, respectively. A magnetic field was not detected in a third
FORS1 measurement reported by \citet{Hubetal09c}. Our re-reduction of all three
observations clearly shows that no field is detected in any of these
three measurements. This star has also been observed three times using the
high-resolution spectropolarimeter ESPaDOnS
 \citep{Wadetal05,Wadetal07b,Aleetal11}. ESPaDOnS data, with
\bz\ uncertainties of about 12~G, show no trace of magnetic field, and
demonstrate quite definitely
that no field anywhere near the level
originally reported is detected in this star. Our conclusion is that the
reported discovery of a magnetic field in HD~139614 is spurious.

The results of a survey with FORS1 of about 50 Herbig AeBe stars were
briefly described by \citet{Wadetal05}. The same paper also reported
first results from the use of the new ESPaDOnS spectropolarimeter at
the CFHT to search for fields in Herbig AeBe stars. From the FORS1
survey, magnetic fields were reported in two stars: HD~72106A (a very
young main sequence star whose companion, HD~72106B, is a Herbig
star), and HD~101412. Although the uncertainties of the measurements
of these two stars have increased in our re-reduction of these data,
both fields are apparently still detected. Since detections are only
at a significance level between 3 and 4\,$\sigma$, FORS1 data by
themselves do not strongly establish the presence of magnetic
fields. In fact, \citet{Wadetal07a} revisited their FORS1 measurement
of HD~72106A, finding that their data could not support a field
detection.  The field of HD~72106A was confirmed by the ESPaDOnS
component of this survey \citep{Wadetal05}, and the star has
subsequently been studied in detail by \citet{Foletal08}.  The
presence of a field in HD~101412 has been fully confirmed by
\citet{Aleetal08a,Aleetal08b,Aleetal09a} with SEMPOL observations, and
by \citet{Hubetal09c} with further FORS1 field measurements at a level
$\ga 6\,\sigma$ (therefore these measurements do not appear in
Table~\ref{Tab_Discoveries}). The field of HD~101412 has been studied
in more detail by \citet{Hubetal11a}.

In the survey reported by \citet{Wadetal05}, the star V380~Ori was
observed with FORS1 without a significant detection. However, the same
paper reports a clear detection of this star (on a different date)
with ESPaDOnS. The magnetic field has been detected at more than the
$3\,\sigma$ level a number of times, and was studied in detail by
\citet{Aleetal09b}.

Further surveys of Herbig AeBe stars using FORS1 were described by
\citet{Hubetal06c} and \citet{Hubetal07a}. Altogether, these papers
reported the discovery of a magnetic field in three Herbig stars:
HD~31648, HD~144432, and HD~144668.  The observations of HD~31648,
presented as a new field discovery by \citet{Hubetal06c}, were
re-reduced by \citet{Hubetal07a}, who reported a final
\bz\ measurement below the $3\,\sigma$ significance.  For HD~144432,
both papers reported a similar field value, but the uncertainty
published by \citet{Hubetal07a} was a factor of almost 2.5 smaller
than estimated by \citet{Hubetal06c}.  Two measurements of HD~144668
published by \citet{Hubetal06c} are marginally significant, but do not
exceed 3\,$\sigma$.  \citet{Hubetal07a} revised both measurements, and
found for this star a new field detection with a $4\,\sigma$
significance. We also note that a survey by \citet{Hubetal09c}, that
we will review later in more detail, reports for HD~144668 a second,
marginally significant detection. Our re-reduction of all these
observations brings the field measurement of HD~31648 by
\citet{Hubetal06c}, and the field measurement of HD~144668 by
\citet{Hubetal07a} to zero within errors. Our re-reduction confirms
the field detection by \citet{Hubetal06c,Hubetal07a} in HD~144432 at
almost 5\,$\sigma$ level, as well as the $\sim 3$\,$\sigma$ detection
from the observations of HD~144668 obtained by
\citet{Hubetal09c}. These same three stars were also observed with
FORS1 \citep[with similar uncertainties to those of][]{Hubetal06c} by
\citet{Wadetal07a}, who did not detect any significant fields. In
addition, magnetic field measurements of HD~31648, HD~144432, and
HD~144668 have been obtained with ESPaDOnS by \citet{Aleetal11}.
These new measurements have error bars between ~40\,G and ~145\,G, and
do not show any strong indication of the presence of fields (although
one field measurement for each star reaches $2\,\sigma$).  Thus, it
appears that the reported detections of HD~144668 and HD~144432 might
possibly be real, but these detections are quite uncertain, and need
substantial new evidence to be convincing.

The FORS1 survey briefly described by \citet{Wadetal05} was analysed
in more detail by \citet{Wadetal07a}. Altogether, 68 field
measurements of 50 stars were carried out. The data were re-reduced by
the authors, who concluded that HD~101412 and BF~Ori probably host
magnetic fields, while HD~36112 and CPD~$-53$~295 might be magnetic,
but that these marginal detections would require confirmation.  After
our new reduction, magnetic fields are still detected in HD~101412 and
CPD~$-53$~295 at a slightly more than 3\,$\sigma$ confidence, while
the measurements of HD~36112 and BF~Ori no longer show field
detection. As discussed above, the detection of HD~101412 was later
fully confirmed. A second measurement of CPD~$-53$~295
\citep[published by][]{Hubetal09c} was reported as field detection but
does not appear significant in our re-reduction of the data.  The only
further data available for HD~36112 or BF~Ori are two observations
with ESPaDOnS or NARVAL for each of the stars, which show no hint of
fields \citep{Aleetal11}.  We conclude that a field might be present
in CPD~$-53$~295, but this result would certainly need further
confirmation. The marginal field detections of HD~36112 and BF~Ori
reported by \citet{Wadetal07a} are probably spurious.

The most recent FORS1 survey of Herbig stars is due to
\citet{Hubetal09c}. Nine field detections at more than the 3$\,\sigma$
level are reported, including two field detections in HD~101412 and
one in HD~190073, two stars previously known to be magnetic Herbig
AeBe stars. New field discoveries are claimed for six stars (on the
basis of one measurement each), although one of these new magnetic
stars is CPD$-53$~295, which was already reported to be possibly
magnetic by \citet{Wadetal07a}. Four of these six detections,
including the measurement of CPD$-53$~295, disappear in our
re-reduction, while two of them remain signficant. As discussed above,
a 3$\,\sigma$ detection of a field in HD~144668, might be real but
requires confirmation. The detection at the $5 - 6\,\sigma$ level of a
field in HD~150193 is even stronger after our re-reduction than originally
published.  However, a single ESPaDOnS observation (with an
uncertainty of about 120~G) detects no field \citep{Aleetal11}.

Finally, we note that our new reduction of the datasets obtained by
\citet{Wadetal07a} yields two field detections in the star HD~97048
(which were not found in the original reductions), at just slightly
over the $3\,\sigma$ level. \citet{Hubetal09c} also has one detection
at about the $4\,\sigma$ level, which by constrast is not retrieved in
our new reduction. It is possible that this star has a weak field of a
few hundred~G, but we do not consider the evidence for this to be very
strong.

In summary, FORS1 observations have successfully been used to discover
the field of one Herbig star, HD 101412, and of the binary companion
of another Herbig star, HD 72660A \citep{Wadetal05}.  FORS1
observations have shown that fields are possibly present, but
certainly not yet definitely established, in HD~97048 and HD~144668
\citep{Hubetal06c}, HD~144332 \citep{Hubetal06c,Hubetal07a},
CPD~$-53$~295 \citep{Wadetal07a,Hubetal09c}, and HD~150193
\citep{Hubetal09c}. All other published FORS1 field discoveries of
Herbig AeBe stars appear to be spurious. Detectable magnetic fields
appear to occur in only a few percent of all Herbig AeBe stars
observed with FORS1. This result is consistent with the 7\,\% found by
\citet{Aleetal11} from the comparable size ESPaDOnS survey.

\subsection{$\beta$ Cephei pulsators and Slowly Pulsating B stars}\label{Sect_Beta_SPB}
The $\beta$~Cep pulsators are early B stars, with spectral types
mostly earlier than B3, and masses in the range
of about 10 to $20\,M_\odot$, which pulsate in at least one radial p mode
with periods in the range of 0.1 to 0.5~d. The Slowly Pulsating B
stars (SPBs) are stars which are B2 or later, have slightly
lower masses than $\beta$~Cep stars, and show non-radial g mode
pulsations with periods of order 1~d. It appears that these pulsations
are excited by the kappa mechanism operating in a deep region of high
iron opacity.

A field was discovered in the prototype $\beta$~Cep star, $\beta$~Cep
= HD~205021 itself \citep{Henetal00,Neinetal01,Donetal01}. The field
of this star appears to be very securely detected (see
Sect.~\ref{Sect_Be_Stars}).

A magnetic field was reported in the $\beta$~Cep star V2052~Oph =
HD~163472 by \citet{Neinetal03b}, again based on MuSiCoS observations.
The measurements presented appear to support the claimed discovery,
but are very close to the limit of detection. The reality of this
field is now clearly confirmed by NARVAL observations \citep{Neinetal11}.

\citet{Neinetal03c} reported the presence of a magnetic field in the
SPB star $\zeta$~Cas = HD~3360 (still based on MuSiCoS data). Although
these observations appear consistent with the presence of a field,
they do not support fully convincing evidence for it. However, the
presence of a magnetic field in $\zeta$~Cas is definitely confirmed by
further (unpublished) observations with MuSiCoS and NARVAL (C.~Neiner,
private communication).

Major surveys of both $\beta$~Cep and SPB pulsators have been carried
out by \citet{Hubetal06a} and \citet{Hubetal09a}. Nearly 70 stars were
observed for magnetic fields using the FORS1 spectropolarimeter (19
known and suspected $\beta$~Cep stars, 50 known and suspected SPBs).
In this survey, fields were reported to have been detected in five
$\beta$~Cep stars and 26 SPB stars. If these results were confirmed,
they would be quite important; such high incidence of detected fields
would strongly suggest that magnetism is intrinsically connected with
the pulsation phenomenon in early B-type stars, as it is in the cool
rapidly oscillating Ap (roAp) stars.

\citet{Hubetal06a} and \citet{Hubetal09a} have reported discovery of a
field in the $\beta$~Cep star $\xi^1$~CMa = HD~46328, based on
altogether 13 detections at approximately the 6 or $7\,\sigma$ level.
The field always appears to be close to $+350$~G. These detections are
further supported by a single field detection by \citet{Siletal09}
using ESPaDOnS at about the $30\,\sigma$ level, by two ESPaDOnS field
measurements reported by \citet{Shuletal11}, and by further ESPaDOnS
observations not yet published in detail \citep{Fouretal11}. The
re-reductions of the FORS1 data (not shown in
Table~\ref{Tab_Discoveries} because the measurements exceed the
$6\,\sigma$ upper limit for inclusion) also confirm the presence of a
field. This discovery appears very robust, and clearly indicates the
capability of FORS1 to detect rather modest magnetic fields in B-type
stars.

The reported field detections in $\beta$~Cep stars in
Table~\ref{Tab_Discoveries} have mostly become insignificant in the
new reductions. In particular, the six reported detections of a field
in HD~16582 all have decreased below the $3\,\sigma$ significance
limit, although one measurement not originally claimed as detection by
\citet{Hubetal09a}, on MJD=54343.259, has risen to the apparently
significant value of $-104 \pm 19$\,G.  A single ESPaDOnS field
measurement detects no trace of a field with \bz\ uncertainty of 10~G
\citep{Siletal09}.  While it is still possible that a field might be
present in HD~16582, this must certainly be regarded at present as --
at best -- extremely uncertain.  Only two published field detections,
one each for HD~74575 and HD~136504, are still significant at the
$3\,\sigma$ level, and each of these measurements is only barely
significant. Three field measurements with ESPaDOnS of HD~74575 show
no trace of a field at the 6~G level \citep{Shuletal11}; we regard
these measurements as clear evidence that no field has been detected
in this star. On the other hand, a field is definitely detected by
\citet{Shuletal11} in two out of two measurements of HD~136504
(=$\epsilon$\,Lupi); these measurements clearly confirm the field
discovered by \citet{Hubetal09a}.  All the remaining reported
detections in $\beta$~Cep stars appear to be the result essentially of
underestimated measurement uncertainties; currently there is no
significant evidence for fields based on these measurements.

Of the more than 40 field detections reported in SPB stars by
\citet{Hubetal06a} and \citet{Hubetal09a}, all but five have decreased
to non-detections in the new reductions.

For HD~28114, the re-reduced field strength is considerably larger
than the published value (and of the opposite sign), but significant
at the $4\,\sigma$ level. Field detection is obtained only on the
higher order Balmer lines, which suggests that our detection is
spurious. A single ESPaDOnS observation finds no trace of a field at
the 30~G level \citep{Siletal09}; it is presently not at all clear
that a field has been detected in HD~28114.

For HD~74195, the published detection is no longer significant, but a
field detection on MJD=54143.072, contradicted by the non-significant
published field value, is now found at slightly more than $3\,\sigma$
significance, so that a field might possibly be present.

In the new reductions, HD~53921 still has one significant field
detection, although with the opposite sign to the published value; the
other two published detections are no longer significant. The star is
a visual binary with 1.4\arcsec separation; in less than optimal
seeing conditions the two components cannot be easily disentangled
(e.g., on MJD=53630.401 and 53631.408 the two spectra were not
distinguishable.)  Note that a field has also been detected in this
star in two HARPSpol measurements recently obtained by one of us
(O.~Kochukhov), so the detection of a field in this star appears to be
secure.

The new reductions of HD~152511 still have two apparently very
significant detections (but we note that for reasons probably related
to bad image quality, our reductions of data obtained on MJD=54609
lead to poor results).

For 16~Peg = HD~208057, the new reductions confirm the original
detection published by \citet{Hubetal06a}, and also indicate a
significant detection of $-195 \pm 60$~G at slightly over the
$3\,\sigma$ significance level for the other measurement of this star
(for which the original published field was not a detection). Note
that a field has also been marginally detected by \citet{Siletal09} in
a single ESPaDOnS measurement of HD~208057; the reality of this field
has also been much more extensively confirmed by \citet{Henretal09}.

In $\la 1$\,\% of the observations, the FORS pipeline failed to reduce
the scientific frames. The dataset of HD~161783 obtained on
MJD=53520.308 is one of these rare cases, and our entry in
Table~\ref{Tab_Discoveries} is an empty space.

We finally note that the field measurement of HD\,215573 obtained on
MJD=53193.321 and published by \citet{Hubetal06a} is obtained from
frames that seem to have reached saturation, even though the ADU count
is just around 50\,000. We found that the CCD gain was set such that
the electron level would reach the full well capacity before ADC
saturation. Accordingly, we have discarded this dataset, and left an
empty space in the entry of Table~\ref{Tab_Discoveries}. The same
problem affects many observations of programme ID 073.D-0466,
dedicated to the observations of SPB and Bp stars.

Based on the re-reductions of the FORS1 data, as well as other
published material, we conclude that the securely detected fields at
present among the $\beta$~Cep pulsators are those of $\beta$~Cep,
V2052~Oph, $\xi^1$~CMa, and HD~136504 (=$\epsilon$~Lupi). Securely
detected fields occur in the SPB stars $\zeta$~Cas = HD~3360,
HD~53921, and 16~Peg = HD~208057.\  $\beta$~Cep star HD~16582, and SPB
stars HD~74195 and HD~152511 may possibly have weak fields, but these
apparent detections still need to be firmly confirmed or rejected. The
great majority of field discoveries announced by \citet{Hubetal06a}
and \citet{Hubetal09a} are spurious.  Based on this conclusion, it
appears that, as in other kinds of upper main sequence stars, magnetic
fields are relatively rare. \citet{GrunWade11} estimate that fields
are present in 16\,\% of pulsating B stars, only marginally more than
the incidence of magnetic fields among A stars. More importantly,
there is currently no significant case for considering that the
pulsation properties of these B pulsators are intrinsically connected
to the presence of weak magnetic fields.

\subsection{$\delta$ Scuti pulsating late A-type stars}
Delta Scuti stars are late A- or early F-type stars, on or near the main
sequence, that pulsate by the kappa mechanism with periods of a few
hours. The pulsations are a mixture of radial and non-radial modes. 

Although $\delta$~Sct stars occur in stars of the same mass and
evolution state as cool magnetic Ap stars, there are no confirmed
magnetic Ap stars that show $\delta$~Sct pulsations. From photometry
with Kepler mission, several possible magnetic Ap $\delta$ Sct stars
have been reported, although the spectral classification of these
stars is still quite uncertain. One definite case of a field in a
pulsation $\gamma$~Dor star has been discovered \citep{Baletal11}. On
the other hand, there does not appear to have been any substantial
survey of known $\delta$~Sct stars for magnetic fields, although
occasionally surveys of bright A- and F-type stars have included one
or two $\delta$~Sct stars such as Altair and $\beta$~Cas
\citep{Lan82,Monetal02}. No field detections have emerged from these
observations, although it would be interesting to carry out a larger
survey of $\delta$~Sct stars.
 
Using FORS1, \citet{Kuretal08} have observed the $\delta$~Sct star
HD~21190, for which an Ap classification was proposed
\citep{Koeetal01}, and have reported an apparently significant field
of $\bz = 47 \pm 13$~G, although they comment that this detection
requires confirmation. Our re-reduction of this measurement increases
the uncertainty only a little, but the result is that there is no
significant detection. We regard this detection as spurious. We also
need to add that a few Ap classifications of known $\delta$~Scuti
stars are inconclusive, since none was supported by a detailed
spectroscopic study. A quick inspection of archival UVES spectra 
allows us to conclude in particular that HD~21190 is not an Ap star.

\subsection{Non-peculiar B-type main sequence and giant stars}
Magnetic fields have begun to be detected in B-type stars that do not
show the obvious chemical peculiarities that signal magnetic Bp or
He-peculiar stars.  The stars $\beta$~Cep, $\zeta$~Cas, and
$\xi^1$~CMa have been discussed above in the context of pulsating B
stars. Fields have also been reported in $\tau$~Sco = HD~149438
\citep{Donetal06b}, HD~36982 = Par~1772 and NU~Ori= HD~37061 
\citep{Petetal08}.

FORS1 surveys including such stars have been reported by
\citet{Hubetal06b}, \citet{McSwain08}, \citet{Hubetal09a} and
\citet{Hubetal09b}. As noted in Table~\ref{Tab_Discoveries},
detections of fields have been reported in several stars. In our new
reductions, the only remaining measurement that shows a significant
field is one of two observations of HD~52089, whose field appears to
be significant at about the $5\,\sigma$ level. \citet{McSwain08}
reported a marginal detection in NGC 3766 MG 45, apparently
contradicted by our estimate of Table~\ref{Tab_Discoveries} obtained
from the full spectrum.  However, from H Balmer lines, we measure
$\bz=-258\pm77$\,G.  In conclusion, a field might be present both
these stars, but would certainly require confirmation. The other
reported detections resulting from these surveys are probably
spurious.

In addition, \citet{Hubetal08a} report a number of magnetic field
measurements of the B0Vp star $\theta$~Car = HD~93030, the brightest
star in the open cluster IC~2602. Five of these measurements are
significant at $\ga 3$\,$\sigma$ level, as listed in
Table~\ref{Tab_Discoveries}, although \citet{Hubetal08a} do not conclude
that the star is magnetic. Because the 26 measurements were all
obtained within about 80 min, in our re-reduction of the field
measurements we have combined all measurements into a single average
value. The average measurement does not show any signficant field, and
we conclude that FORS data do not reveal presence of a magnetic field.

\subsection{O-type stars}
O-type stars are relatively difficult targets for magnetic field
measurements, because they have few strong spectral lines, high
projected rotational velocities ($\sim 100$~\kms), and often fairly
strong emission filling in some of the photospheric absorption lines.
The first detections came when a field was detected in
$\theta^1$~Ori~C, the brightest star in the Trapezium, by
\citet{Donetal02}, and in the Of?p star HD~191612 by
\citet{Donetal06a}, using high-resolution spectropolarimetery.

A field of order 100~G has been reported in the O9.7 supergiant
  $\zeta$~Ori~A = HD~37742 \citep{Bouretal08}, but the field detection
  is near the limit of detectability and needs  confirmation.

\citet{Hubetal08b} carried out the first large survey of O-type stars
with FORS1, and reported field measurements of 13 O stars
\citep[including one star previously reported by][]{Hubetal07b}. Field
discoveries were claimed for five of these stars, as listed in
Table~\ref{Tab_Discoveries}. When reduced with our tools, only one of
these stars still shows a significant magnetic field, the O6.5f?p star
HD~148937. The field of this star has more recently been detected in
three more observations with FORS2 by \citet{Hubetal11b}. Furthermore,
the field has been confirmed by \citet{Wadetal11} from observations
obtained by the MiMeS project. This discovery is clearly real, but it
appears that the field of this star, always about $-250$~G, is close
to the limit for reliable field detection by FORS1 in such stars.

The other O-type stars for which fields are reported by
\citet{Hubetal08b} appear to be spurious detections.

\citet{Hubetal11c} have reported the detection of a magnetic field at
just over the $3\,\sigma$ level in the runaway O-type star $\zeta$~Oph =
HD~149757. Our re-reduced measurement of this star no longer shows a
significant field. This detection is probably spurious.

\subsection{X-ray binary star systems}
Under this heading, we consider two different types of possibly
magnetic stars: nearby A-type stars identified with X-ray sources in the
ROSAT catalogue, and the brilliant and distant X-ray source Cyg~X-1.

A-type stars are not expected to emit detectable levels of X-ray flux. They
are not hot enough to have the strong radiatively driven winds whose
instabilities or collision regions are strong X-ray generators, and
the A-type stars do not have the deep convection zones that lead to stellar
activity and essentially universal coronal X-ray emission. However,
about 15\% of A-type stars are associated with detected X-ray sources. Some
of these X-ray sources are low-mass, active companions, but it is not
clear that in all cases this is an adequate explanation.

\citet{Schetal08} have carried out an extensive survey with FORS1 for
magnetic fields in X-ray emitting A-type stars in which the companion
hypothesis seems doubtful. This survey has led to three reported
discoveries of fields in A-type stars, as listed in
Table~\ref{Tab_Discoveries}. In the cases of HD~147084 and HD~159312,
the new reductions show no significant field, and these detections are
probably spurious. The other detection is in HD~148898, which is known
to be a magnetic Ap star. Although no field has previously been
convincingly detected in this star \citep{BorLan80}, it is expected
\citep{Auretal07} to have a weak magnetic field. The field reported by
\citet{Schetal08} is still present at almost the $3\,\sigma$ level in
our new reductions, and may be correctly detected by this observation.

The system Cyg~X-1 = HD~226868 is a binary composed of a black hole
orbiting with an O9.7Iab supergiant, which is losing mass onto the
black hole. FORS1 has been used to search for the presence of a
magnetic field in the optical primary star of this system by
\citet{Karetal09,Karetal10}. They report 13 field measurements spread
over 1.5 months. Three of their measurements show apparently
significant detections. In our new reductions, two of these three
measurements are still significant at about $5\,\sigma$ level. Two
other measurements, not showing a $3\,\sigma$ detection in the
original reductions, reveal a $\ga 3$\,sigma field detection in our
new reductions. The field apparently detected in this star is very
near the reliable detection limit for FORS1. We also note that due to
the target declination, all observations were obtained with the
telescope at a large zenith distance ($\sim 60\degr$), and that in the
course of the 1.1\,h long observations, the instrument was rotating by
$\sim 20\degr$. Should flexures play a role in the magnetic field
determinations obtained with FORS, they certainly have an impact on
these observations. Finally, we must remark that we measure an
unusually high fraction of circular polarization in the continuum
(about $-0.4$\,\% at 3600\,\AA, linearly increasing up to about
$-0.3$\,\% at 5200\,\AA). While this signal is possibly real, it could
potentially be due to cross-talk from linear polarization \citep[see
Sect.~7.4 of][]{Bagetal09}. It is clear that further observations of
Cyg~X-1, both in linear and circular polarization, would be of
considerable interest.

\subsection{Hot subdwarfs}
Detections of kG magnetic fields in two sdO stars, 
Feige 66 = BD +25$^\circ$ 2534, and BD +75$^\circ$ 325,
were reported by \citet{Elki96}.

Magnetic fields were measured using FORS1 in six hot subdwarfs by
\cite{Otoetal05}, and significant detections were claimed for all six
stars, as listed in Table~\ref{Tab_Discoveries}.  Four of these stars
are B subdwarfs, probably stars that are essentially in the horizontal
branch stage of evolution but have somehow lost virtually all their
hydrogen envelopes. The other two, CD~$-31$~4800 and LSE~153, are O
subdwarfs, probably stars on their way from being central stars of
planetary nebulae to being white dwarfs.

In our new reductions, the uncertainties of the \bz\ measurements are
slightly reduced compared to the values originally published, but no
detections at even the $2\,\sigma$ level are found. This survey
provides no significant evidence for the presence of magnetic fields
in such stars at the level of 1~kG. Furthermore, new ESPaDOnS
measurements of HD~76431, one of the sdB stars reported to host a kG
field by \citet{Otoetal05}, find no indication of any field above a
level of order 100~G \citep{Petietal11}. \citet{Petietal11} have also
shown that the FORS1 field measurement of HD~76431 does not support any
significant field detection. Finally, they also report no detection in
Feige 66, for which \citet{Elki96} claimed detection of a kG field.

\citet{Valetal06} have carried out a search for weak (kG) magnetic
fields in a number of white dwarfs, and the sdO star WD~1036+433,
using the spectropolarimeter on the Russian 6-m telescope of the
Special Astrophysical Observatory. Of the three observations of this
subdwarf, one reveals a field of $9.6 \pm 2.6$~kG, at the $3.7\,\sigma$
level, which is suggestive of the presence of a field, but not yet
conclusive. 

At present there is at most marginal evidence for fields of a few kG
in subdwarfs, and it appears that most of the reported detections are
spurious. However, the total sample studied is very small. This is a
category of star that would clearly benefit from a larger survey with
FORS.

\subsection{Central stars of planetary nebulae}
\citet{Joretal05} carried out six field measurements of four central
stars of planetary nebulae (CSPN), and concluded that at least two of
these stars have highly significant fields in the kG range. This
result supports the idea that such fields have an important shaping
effect on the the planetary nebulae themselves. It also suggests that
important loss of magnetic flux must occur between this evolution
state and that of white dwarf, since white dwarfs with as much
magnetic flux as that inferred for the CSPN are relatively
rare. These results were called into question recently by
\citet{Leoetal11}, who obtained one new measurement of each of the
CSPNs NGC~1360 and LSS~1362, detecting no significant field in either
star, and who re-reduced the older data of \citet{Joretal05}, also
finding no significant fields.

Our new reductions are completely consistent with the re-reductions
carried out by \citet{Leoetal11}. None of the original field
measurements by \citet{Joretal05} reveals a significant field. As
remarked by \citet{Leoetal11}, there is now no significant evidence
for coherent magnetic fields in the central stars of planetary
nebulae, and the current best upper limits on \bz\ are at roughly the
1 to 2~kG level.

\subsection{(Weak-field) magnetic white dwarfs}
It has been known for 40 years that strong magnetic fields occur in a
small fraction of white dwarf stars. The fields can be as large as
$\sim 300$~MG, and as small as tens of kG. In recent years, there has
been considerable interest in establishing how the low field end of
the distribution of field strengths found in the sample of magnetic
white dwarfs behaves: do most or all white dwarfs have weak (a few kG)
fields, or are weak fields as rare as the stronger fields
are, occurring in only a few percent of white dwarfs?

A first effort to answer this question was made by \citet{SchSmi95},
who surveyed nearly 170 DA white dwarfs (i.e. white dwarfs with strong
Balmer line spectra) using the Zeeman effect in the H
lines. \citet{SchSmi95} found two white dwarfs with fields of order
$10^2$~kG. More recently, a smaller survey was carried out by
\citet{Fabetal03} and \citet{Valetal06}, using the Russian 6-m
telescope, which has reported several field detections. The fields
detected range in size from a few kG to a few tens of kG. 

FORS1 has made a substantial contribution to this search.
\citet{Aznetal04} detected kG fields in three white dwarfs.
\citet{Joretal07} have continued the survey, and identified three more
stars that they consider probably weakly magnetic (although no single
field measurement reached the $3\,\sigma$ level of significance). Our
re-reduced data for the three magnetic white dwarfs reported by
\citet{Aznetal04} still show field values that consistently differ
from zero by between 3.3 and $6.2\,\sigma$. Although none of these
measurements is completely outside of the range of spurious
detections, the fact that significant fields are detected in two
measurements each of WD~0446-789 and WD~2359-434, and that the single
measurement of WD~1105-048 is significant at the $5.3\,\sigma$ level,
suggests that all three field detections may well be correct. However,
confirming observations are required. In contrast, the (less
significant) field measurements reported by \citet{Joretal07} all have
diminished to an insignificant level in the new reductions. At present
there is no strong evidence for fields in any of these stars.

Note that the sign of the field values has been reversed to conform to
the convention used in magnetic main sequence stars.

Since the fields apparently detected with FORS1 are among the very
smallest fields known in white dwarfs, it is clear that the
instrument is a very powerful tool for this kind of work, and that
further observations would be of considerable value.

\section{Conclusions}
We have investigated the robustness of magnetic field detections
obtained with the FORS1 instrument of the ESO VLT. To carry out our
study, we have developed a semi-automatic data reduction procedure and
applied it to the entire FORS1 archive of circular spectropolarimetric
data.

We have performed sophisticated Monte Carlo simulations (based
on the repetition of the entire data reduction process after altering the
original raw frames to simulate photon noise), and concluded that our
estimate of the field error bar correctly accounts for the photon noise.

We have performed all our field measurements adopting several
different and reasonable data reduction algorithms, and concluded that
the taking certain (somewhat arbitrary) choices during the process of
data reduction contributes to the budget of the measurement error in a
significant way.

We have determined the magnetic field from the null profiles, and
studied its distribution. Our conclusion is that photon-noise is not
the only source of uncertainties.  Among the possible causes of
non-Poisson noise, we find that small instrument flexures, negligible
in most of the other instrument applications, set a lower limit to the
accuracy of the field measurements, and may be responsible for 
some of the field detections in the null profiles.

Overall, non-statistical noise cannot be characterised in
detail, but as a rough guide we estimate that actual error bars
may be some 50\,\% higher than estimated from photon-noise,
especially for high signal-to-noise ratio measurements.

We have performed a detailed comparison with the results previously
published in the literature. Even without considering the impact of
non-photon noise, we are unable to confirm many of the previously
published new discoveries that were obtained at a 3 to 6\,$\sigma$
level, and we have discussed how the incidence of the magnetic field
in various classes of stars should be revised. The fact that our
analysis of FORS1 data does not allow us to confirm many of the recent
discoveries, leads us to conclude that there is no contradiction
between the results obtained with FORS1 and those obtained with
high-resolution spectropolarimeters.

FORS is a low resolution spectropolarimeter perfectly suitable for
the detection of magnetic fields with $\ga 300$\,G strength. Fields
weaker than about 100\,G are probably out of its reach. In general, we
argue that any field detection reported at a less than 5-6\,$\sigma$
level should be corroborated by repeated observations, and, if
possible, cross-checked with other instruments. While it cannot reach
the same accuracy of a high-resolution instrument sitting in a
thermally and mechanically isolated bench (like ESPaDOnS and
HARPSpol), FORS remains an invaluable instrument for the detection of
magnetic fields in faint and/or rapidly rotating stars.

\begin{acknowledgements}
This work is based on data obtained from the ESO Science Archive
Facility, under the following programmes: 
069.D-0210, 070.D-0259, 071.D-0308, 270.D-5023, 072.C-0447, 072.D-0089,
072.D-0290, 072.D-0377, 272.C-5063, 073.D-0356, 073.D-0466, 073.D-0464,
074.C-0442, 074.C-0463, 075.D-0295, 075.D-0432, 075.D-0507, 077.D-0406,
077.D-0556, 078.D-0140, 278.D-5056, 079.D-0241, 079.D-0549, 080.D-0383,
081.C-0410, 381.D-0138.\\
JDL has been supported by the Natural Sciences and Engineering
Research Council of Canada. \\
We thank E.~Alecian and C.~Nenier,
for many useful discussions and suggestions, and the referee, P.~Petit,
for his constructive comments.\\
SB would like to dedicate his contribution to this paper to
the memory of Carlo Izzo, dearest friend, and precious collaborator.
We acknowledge the crucial importance for this paper provided by
his work at ESO, and by numerous useful discussions. Carlo Izzo also
kindly provided us with some of the scripts used for data organisation.
\end{acknowledgements}

\Online
\begin{center}
\begin{longtable}{llcr@{$\pm$\,}llr@{$\pm$\,}lc}
\caption{\label{Tab_Discoveries} List of published ``marginal'' field detections}\\
\hline\hline
Class &Star & MJD  &\multicolumn{2}{c}{\bz\ (G)}   & Reference &\multicolumn{2}{c}{\bz\ (G)}     & Field      \\
      &Name &      &\multicolumn{2}{c}{(published)}&           &\multicolumn{2}{c}{(this work)}  & detected?  \\
\hline
\endhead
\hline
\endfoot
\hline\hline
\endlastfoot
\multicolumn{2}{l}{\bf HgMn and PGa stars}      \\[1mm]
 &HD 358     &  52910.092  &  $-261 $&$ 73$  & HNS06 & $-296 $&$ 121$ & N  \\ 
 &           &  53629.286  &  $ -73 $&$ 20$  & HNS06 & $  23 $&$  33$ &    \\ 
 &           &  53638.205  &  $-108 $&$ 23$  & HNS06 & $ -36 $&$  37$ &    \\ 
 &HD 19400   &  52852.371  &  $ 217 $&$ 65$  & HNS06 & $ 124 $&$  85$ & N  \\
 &HD 65949   &  53002.082  &  $-290 $&$ 62$  & HNS06 & $ -42 $&$  79$ & N  \\ 
 &HD 65950   &  53002.067  &  $-179 $&$ 53$  & HNS06 & $  38 $&$  74$ & N  \\ 
 &HD 175640  &  52901.037  &  $ 207 $&$ 65$  & HNS06 & $  31 $&$  64$ & N \\

\multicolumn{2}{l}{\bf Be stars}\\[1mm]
 &HD 56014   &  53511.976  &  $-146 $&$ 32$  & HYP07 & $-114 $&$  48$ & N \\ 
 &HD 62367   &  54549.096  &  $  99 $&$ 32$ & HSSY09 & $  36 $&$  51$ & N   \\
 &HD 120324  &  53869.296  &  $ -80 $&$ 24$ & HSSY09 & $  78 $&$ 111$ & N   \\
 &HD 148184  &  53532.224  &  $  83 $&$ 21$ & HYP07  & $ 113 $&$ 120$ & N   \\
 &           &  53862.380  &  $ 136 $&$ 16$ & HYP07  & $ -593$&$ 522$ &     \\
 &HD 181615/6&  53520.415  &  $  38 $&$ 10$ & HYP07  & $ 28  $&$  28$ & N   \\
 &           &  54333.020  &  $ -78 $&$  8$ & HSSY09 & $ 15  $&$  11$ &     \\
 &           &  54343.098  &  $ -73 $&$  9$ & HSSY09 & $ 56  $&$  10$ &     \\
 &HD 209409  &  53955.185  &  $  85 $&$ 28$ & HSSY09 & $-42  $&$  46$ & N   \\
 &HD 224686  &  54432.065  &  $  74 $&$ 24$ & HSSY09 &\multicolumn{2}{c}{}& N   \\
 &NGC 3766 MG 47&54549.021 &  $-234 $&$ 69$ & McS08  & $-61  $&$  60$ & N   \\
 &           &             &  $-134 $&$ 42$ & HSSY09 &\multicolumn{2}{c}{}&    \\ 
 &NGC 3766 MG 200&54550.375&  $ -12 $&$ 42$ & McS08  & $ -1  $&$  56$ & N   \\
 &            &            &  $ 128 $&$ 40$ & HSSY09 &\multicolumn{2}{c}{}&     \\ \\

\multicolumn{2}{l}{\bf Herbig AeBe stars} \\[1mm]
 &HD 31648   &  53296.355  &  $  87 $&$ 22$ & HYS06  & $ -33 $&$  50$ & N  \\ 
 &HD 36112   &  53331.210  &  $-149 $&$ 41$ & WBD07  & $ -73 $&$  67$ & N  \\ 
 &HD 72106A  &  53332.296  &  $ 195 $&$ 45$ & WDB05  & $ 206 $&$  58$ & Y   \\
 &           &             &  $  94 $&$ 55$ & WBD07  &\multicolumn{2}{c}{}& \\
 &HD 97048   &  54609.138  &  $ 164 $&$ 42$ & HSSG09 & $ 102 $&$  62$ & P  \\
 &HD 100546  &  54610.044  &  $  89 $&$ 26$ & HSSG09 & $ -43 $&$  40$ & N  \\
 &HD 101412  &  53062.299  &  $ 430 $&$ 75$ & WDB05  & $ 471 $&$ 136$ & Y  \\
 &           &             &  $ 446 $&$106$ & WBD07  &\multicolumn{2}{c}{}&   \\
 &HD 135344B &  54610.144  &  $ -38 $&$ 11$ & HSSG09 & $ -14 $&$  17$ & N  \\
 &HD 139614  &  52904.040  &  $-450 $&$ 93$ & HSY04  & $ -84 $&$  65$ & N  \\ 
 &           &             &  $-150 $&$ 50$ & WDB05  &\multicolumn{2}{c}{}&   \\ 
 &           &             &  $-450 $&$ 93$ & HYS06  &\multicolumn{2}{c}{}&   \\ 
 &           &             &  $-112 $&$ 36$ & HPY07  &\multicolumn{2}{c}{}&   \\ 
 &           &  53405.373  &  $-116 $&$ 34$ & HYS06  & $   8 $&$  25$ &   \\
 &           &             &  $ -93 $&$ 14$ & HPY07  &\multicolumn{2}{c}{}&   \\ 
 &HD 144432  &  53447.352  &  $-119 $&$ 38$ & HYS06  & $-108 $&$  22$ & P  \\
 &           &             &  $-111 $&$ 16$ & HPY07  &\multicolumn{2}{c}{} &   \\
 &HD 144668  &  52901.007  &  $ 166 $&$ 40$ & HPY07  & $ -90 $&$  85$ & P  \\  
 &           &  54610.238  &  $ -62 $&$ 18$ & HSSG09 & $-108 $&$  32$ &    \\
 &HD 150193  &  54609.093  &  $-144 $&$ 32$ & HSSG09 & $-238 $&$  40$ & P  \\
 &HD 176386  &  54610.272  &  $-119 $&$ 33$ & HSSG09 & $ -81 $&$  41$ & N  \\
 &HD 190073  &  54609.410  &  $ 104 $&$ 19$ & HSSG09 & $ -47 $&$  57$ & Y  \\
 &CPD $-$53 295&53330.085  &  $ 129 $&$ 32$ & WBD07  & $ 132 $&$  36$ & P  \\ 
 &           &  54610.399  &  $ 103 $&$ 29$ & HSSG09 & $  -5 $&$  32$ &    \\
 &BF Ori     &  53330.277  &  $-144 $&$ 21$ & WBD07  & $-102 $&$  45$ & N  \\ \\

\multicolumn{2}{l}{\bf $\beta$ Cep stars}\\[1mm]
 &HD 16582   &  54109.066  &  $ -88 $&$ 24$ & HBD09  & $-171 $&$ 86$  & P   \\ 
 &           &  54344.200  &  $ -40 $&$ 12$ & HBD09  & $  -1 $&$ 22$  &     \\ 
 &           &  54344.264  &  $ -44 $&$ 12$ & HBD09  & $  47 $&$ 23$  &     \\ 
 &           &  54345.203  &  $ -42 $&$ 12$ & HBD09  & $ -66 $&$ 24$  &     \\ 
 &           &  54345.246  &  $ -49 $&$ 13$ & HBD09  & $  -3 $&$ 21$  &     \\ 
 &           &  54345.293  &  $ -31 $&$ 11$ & HBD09  & $  55 $&$ 20$  &     \\ 
 &HD 50707   &  54107.318  &  $ 163 $&$ 52$ & HBD09  & $  54 $&$ 70$  & N   \\ 
 &           &  54345.372  &  $ 149 $&$ 19$ & HBD09  & $  67 $&$ 27$  &     \\ 
 &HD 74575   &  54082.341  &  $ 142 $&$ 48$ & HBD09  & $-110 $&$ 67$  & N   \\ 
 &           &  54109.150  &  $ 132 $&$ 50$ & HBD09  & $-316 $&$ 76$  &     \\ 
 &HD 136504  &  54344.998  &  $-156 $&$ 34$ & HBD09  & $-132 $&$ 38$  & Y   \\ 
 &HD 180642  &  54344.084  &  $ 166 $&$ 41$ & HBD09  & $  73 $&$ 38$  & N   \\ \\

\multicolumn{2}{l}{\bf SPB stars} \\[1mm]
 &HD 3379    &  53244.402  &  $ 272 $&$ 57$ & HBS06  & $  84 $&$ 50$  & N  \\ 
 &           &  53245.214  &  $ 231 $&$ 47$ & HBS06  & $ -41 $&$ 38$  &    \\
 &           &  54344.233  &  $ 117 $&$ 34$ & HBD09  & $  36 $&$ 36$  &    \\ 
 &           &  54345.189  &  $ 155 $&$ 42$ & HBD09  & $ -14 $&$ 40$  &    \\ 
 &HD 11462   &  54344.248  &  $ 161 $&$ 46$ & HBD09  & $  23 $&$ 45$  & N  \\ 
 &HD 24587   &  54086.175  &  $-353 $&$ 82$ & HBD09  & $-140 $&$ 71$  & N  \\ 
 &HD 25558   &  54345.264  &  $ 105 $&$ 34$ & HBD09  & $  37 $&$ 40$  & N  \\ 
 &HD 28114   &  54106.091  &  $ 107 $&$ 33$ & HBD09  & $-450 $&$104$  & N  \\ 
 &HD 28475   &  54345.312  &  $ 160 $&$ 48$ & HBD09  & $   7 $&$ 41$  & N  \\ 
 &HD 40494   &  54343.426  &  $  94 $&$ 28$ & HBD09  & $   9 $&$ 29$  & N  \\ 
 &HD 45284   &  53252.365  &  $ 245 $&$ 63$ & HBS06  & $  28 $&$ 47$  & N  \\ 
 &HD 53921   &  52999.137  &  $-294 $&$ 63$ & HNS06, 
                                              HBS06  & $ 486 $&$ 89$  & Y  \\ 
 &           &  53630.401  &  $ 151 $&$ 29$ & HBS06  & $ 153 $&$ 158$ &    \\ 
 &           &  53631.408  &  $ 151 $&$ 21$ & HBS06  & $ 154 $&$ 123$ &    \\ 
 &HD 74195   &  53138.972  &  $ 310 $&$ 98$ & HBS06  & $  65 $&$  83$ & P  \\
 &HD 74560   &  53002.141  &  $-199 $&$ 61$ & HNS06, 
                                              HBS06  & $ 259 $&$ 95$ &  N  \\ 
 &           &  54108.348  &  $-198 $&$ 55$ & HBD09  & $ -68 $&$ 60$ &     \\ 
 &HD 85953   &  53002.152  &  $-131 $&$ 42$ & HNS06, 
                                              HBS06  & $  14 $&$ 55$ &  N  \\ 
 &           &  54156.096  &  $  97 $&$ 29$ & HBD09  & $  12 $&$ 39$ &     \\ 
 &HD 140873  &  53151.192  &  $ 286 $&$ 60$ & HBS06  & $   4 $&$ 56$ &  N  \\ 
 &           &  54344.011  &  $  99 $&$ 31$ & HBD09  & $ -29 $&$ 36$ &     \\
 &HD 152511  &  54344.116  &  $ 649 $&$ 43$ & HBD09  & $ 595 $&$  51$ & P  \\
 &           &  54608.158  &  $ 141 $&$ 26$ & HBD09  & $  66 $&$  32$ &    \\
 &           &  54609.433  &  $ 440 $&$ 39$ & HBD09  & $ 117 $&$ 570$ &    \\
 &           &  54610.223  &  $ 158 $&$ 28$ & HBD09  & $ 118 $&$  33$ &    \\
 &HD 152635  &  54344.041  &  $-149 $&$ 36$ & HBD09  & $ -88 $&$  30$ & N  \\
 &HD 160124  &  53151.259  &  $ 456 $&$ 60$ & HBS06  & $ -43 $&$  57$ & N  \\
 &HD 161783  &  53151.281  &  $ 376 $&$ 63$ & HBS06  & $  40 $&$  51$ & N  \\
 &           &  53520.308  &  $-113 $&$ 32$ & HBS06  &\multicolumn{2}{c}{} &   \\ 
 &HD 163254  &  54344.068  &  $ 155 $&$ 49$ & HBD09  & $ -49 $&$  53$ & N  \\
 &HD 169467  &  54345.164  &  $-182 $&$ 41$ & HBD09  & $ -74 $&$  39$ & N  \\
 &HD 169820  &  53151.312  &  $ 239 $&$ 70$ & HBS06  & $  33 $&$  60$ & N  \\
 &HD 179588  &  54343.134  &  $ 158 $&$ 41$ & HBD09  & $  51 $&$  41$ & N  \\
 &HD 181558  &  53227.184  &  $ 236 $&$ 75$ & HBS06  & $  74 $&$  63$ & N  \\
 &           &  53275.143  &  $-336 $&$ 63$ & HBS06  & $  46 $&$  55$ &    \\
 &           &  54344.167  &  $-104 $&$ 32$ & HBD09  & $ -90 $&$  36$ &    \\
 &HD 183133  &  54344.179  &  $ 152 $&$ 38$ & HBD09  & $  81 $&$  44$ & N  \\
 &HD 205879  &  54343.226  &  $ 150 $&$ 40$ & HBD09  & $  98 $&$  49$ & N  \\
 &HD 208057  &  53597.166  &  $-156 $&$ 31$ & HBS06  & $ -124$&$  36$ & Y  \\
 &HD 215573  &  52900.080  &  $ 165 $&$ 53$ & HNS06, 
                                              HBS06  & $ 146 $&$  70$ & N  \\
 &           &  53191.222  &  $ 180 $&$ 54$ & HBS06  & $  -4 $&$  46$ &    \\
 &           &  53193.321  &  $-320 $&$ 90$ & HBS06  &\multicolumn{2}{c}{}&    \\ \\

\multicolumn{2}{l}{\bf $\mathbf{\delta}$ Sct stars}  \\[1mm]
 &HD 21190   &  54343.280  &  $47   $&$ 13$ & KHG08  & $  10 $&$  17$ & N  \\ \\

\multicolumn{2}{l}{\bf Normal B stars} \\[1mm]
 &HD 52089     &  54046.339 &  $-200 $&$ 48$ & HBD09  & $ -127$&$  60$ & P  \\
 &             &  54343.389 &  $-129 $&$ 34$ & HBD09  & $ -196$&$  37$ &    \\
 &$\theta$ Car &  54181.025 & $ 266 $&$ 52$ & HBM08  &\multicolumn{2}{c}{}& N  \\ 
 &             &  54181.035 & $-458 $&$118$ & HBM08  &\multicolumn{2}{c}{}&    \\ 
 &             &  54181.037 & $-394 $&$132$ & HBM08  & $   8 $&$  23$ &    \\ 
 &             &  54181.041 & $-210 $&$ 52$ & HBM08  &\multicolumn{2}{c}{}&    \\
 &             &  54181.053 & $-394 $&$125$ & HBM08  &\multicolumn{2}{c}{}&    \\ \\
 &NGC 3766 MG 45& 54550.067 & $-185 $&$ 53$ & McS08  & $-61  $&$  54$ & P   \\
 &             &            & $-123 $&$ 40$ & HSSY09 &\multicolumn{2}{c}{}&    \\ 
% &NGC 3766 MG 94& 54550.327 & $ 276 $&$ 55$ & McS08  & $  141$&$  72$ & N  \\
% &             &            & $ 294 $&$ 53$ & HSSY09 &\multicolumn{2}{c}{}&  \\  
 &NGC 3766 MG 111&54549.020 & $  54 $&$ 33$ & McS08  & $  -11$&$  40$ & N  \\ 
 &             &            & $ 112 $&$ 34$ & HSSY09 &\multicolumn{2}{c}{}&    \\ 
 &NGC 3766 MG 176&54550.016 & $   3 $&$ 31$ & McS08  & $   -4$&$  39$ & N  \\ 
 &             &            & $  89 $&$ 28$ & HSSY09 &\multicolumn{2}{c}{} \\ 
 &HD 153716    &  54344.057 & $ 124 $&$ 41$ & HBD09  & $  -83$&$  50$ & N  \\
 &HD 179761    &  52822.280 & $-267 $&$ 68$ & HNS06  & $ -222$&$  84$ & N  \\ \\

\multicolumn{2}{l}{\bf Normal O stars} \\[1mm]
 &HD 36879   &  54345.389  &  $ 180 $&$ 52$ & HSS08  & $  33 $&$  58$ & N  \\
 &HD 148937  &  54550.416  &  $-276 $&$ 88$ & HSS08  & $-242 $&$  83$ & Y  \\
 &HD 149757  &  54609.340  &  $ 145 $&$ 45$ & HOS11  & $ 129 $&$  58$ & N  \\
 &HD 152408  &  53556.216  &  $ -89 $&$ 29$ & HSS08  & $ -51 $&$ 114$ & N  \\
 &HD 155806  &  53531.775  &  $-115 $&$ 37$ & HYP07  & $  70 $&$  54$ & N  \\ 
 &HD 164794  &  53594.119  &  $ 211 $&$ 57$ & HSS08  & $ 169 $&$  76$ & N  \\ \\

\multicolumn{2}{l}{\bf X-ray stars} \\[1mm]
 &HD 147084  &  53975.968  &  $ -48 $&$ 15$ & SHS08  & $  46 $&$  25$ & N  \\ 
 &HD 148898  &  52763.349  &  $ 122 $&$ 29$ & SHS08  & $ 184 $&$  63$ & P  \\
 &HD 159312  &  53976.178  &  $ 241 $&$ 80$ & SHS08  & $ 127 $&$  78$ & N  \\
 &HD 226868  &  54291.268  &  $ 101 $&$ 18$ & KBH10  & $ 163 $&$  35$ & P  \\
 &           &  54664.194  &  $  80 $&$ 23$ & KBH10  & $ 260 $&$  44$ &    \\
 &           &  54678.178  &  $ 128 $&$ 21$ & KBH10  & $  69 $&$  34$ &    \\ \\

\multicolumn{2}{l}{\bf Hot subdwarfs}  \\[1mm]
 &UVO 0512$-$08& 53058.025   & $-1306 $&$ 161$ & OJF05 & $  155 $&$ 163$ & N \\
 &CPD $-$64 481& 53058.069   & $ -885 $&$ 207$ & OJF05 & $  -72 $&$ 247$ & N \\
 &PG 0909+276 &  53058.139   & $-1448 $&$ 222$ & OJF05 & $ -258 $&$ 220$ & N \\
 &CD $-$31 4800& 53058.215   & $-1050 $&$ 161$ & OJF05 & $  -34 $&$ 130$ & N \\
 &HD 76431    &  53058.255   & $-1096 $&$  91$ & OJF05 & $   44 $&$  88$ & N \\
 &LSE 153     &  53058.347   & $-1128 $&$ 212$ & OJF05 & $  -44 $&$ 156$ & N \\ \\

\multicolumn{3}{l}{\bf Central stars of planetary nebulae} \\[1mm]
 &NGC 1360   &  52946.291    & $-1343 $&$ 259$  & JWO05 & $  313 $&$ 286$ & N \\
 &           &  52988.235    & $ 1708 $&$ 257$  & JWO05 & $  374 $&$ 262$ &   \\
 &           &  52989.060    & $ 2832 $&$ 269$  & JWO05 & $  564 $&$ 308$ &   \\
 &EGB 5      &  52988.347    & $ 1992 $&$ 562$  & JWO05 & $-1009 $&$ 773$ & N \\
 &LSS 1362   &  52989.302    & $ 1891 $&$ 371$  & JWO05 & $  -11 $&$ 394$ & N \\
 &Abell 36   &  53031.287    & $ 1169 $&$ 466$  & JWO05 & $ 1048 $&$ 379$ & N \\ \\

\multicolumn{2}{l}{\bf Weak-field white dwarfs} \\[1mm]
 &WD 0446$-$789& 52609.229   & $ 3115 $&$ 763$ & AJN04 & $-2765 $&$ 825$ & Y  \\ 
 &            &  52668.087   & $ 6321 $&$ 929$ & AJN04 & $-5355 $&$ 974$ &    \\
 &WD 1105$-$048& 52669.305   & $-3959 $&$ 710$ & AJN04 & $ 3587 $&$ 673$ & P  \\
 &WD 1620$-$391& 53136.301   & $-1580 $&$ 475$ & JAN07 & $   35 $&$ 306$ & N  \\ 
 &WD 2007$-$303& 53132.382   & $-1093 $&$ 453$ & JAN07 & $  563 $&$ 365$ & N  \\ 
 &WD 2039$-$202& 53167.393   & $-1297 $&$ 512$ & JAN07 & $  781 $&$ 598$ & N  \\ 
 &WD 2359$-$434& 52583.025   & $-4504 $&$ 958$ & AJN04 & $ 4248 $&$ 878$ & Y  \\
 &            &  52608.056   & $-2919 $&$ 526$ & AJN04 & $ 3211 $&$ 524$ &    \\

\end{longtable}
\end{center}

\noindent
Key to reference:
AJN04:  \citet{Aznetal04};\ 
JWO05:  \citet{Joretal05};\ 
JAN07:  \citet{Joretal07};\ 
HSY04:  \citet{Hubetal04a};\ 
HBS06:  \citet{Hubetal06a};\         
HNS06:  \citet{Hubetal06b};\         
HYS06:  \citet{Hubetal06c};\         
HPY07:  \citet{Hubetal07a};\         
HYP07:  \citet{Hubetal07b};\         
HBM08:  \citet{Hubetal08a};\         
HSS08:  \citet{Hubetal08b};\         
HBD09:  \citet{Hubetal09a};\         
HSSY09: \citet{Hubetal09b};\ 
HSSG09: \citet{Hubetal09c};\         
KBH10:  \citet{Karetal10};\ 
KHG08:  \citet{Kuretal08};\ 
McS08:  \citet{McSwain08};\ 
OJF05:  \citet{Otoetal05};\ 
SHS08:  \citet{Schetal08};\ 
WDB05:  \citet{Wadetal05};\ 
WAB06:  \citet{Wadetal06};\ 
WBD07:  \citet{Wadetal07a}.

\end{document}